\documentclass[reprint,floatfix]{revtex4-1}

\usepackage{epsf}
\usepackage{epsfig}
\usepackage{color}
\usepackage{verbatim}
\usepackage{amsopn}
\usepackage{appendix}
\usepackage{amsmath}
\usepackage{graphicx}
\usepackage{dsfont}
\DeclareMathOperator*{\sgn}{sign}
\DeclareMathOperator*{\im}{Im}
\DeclareMathOperator*{\re}{Re}

\DeclareMathOperator*{\Shi}{Shi}
\DeclareMathOperator*{\Chi}{Chi}

\begin{document}

\unitlength = 1mm

\title{From non equilibrium quantum Brownian motion to impurity dynamics in one-dimensional quantum liquids}
\author{Julius Bonart}
\affiliation{$^1$Laboratoire de Physique Th\'eorique et Hautes Energies,
Universit\'e Pierre et Marie Curie -- Paris VI, 
4 Place Jussieu, 
75252 Paris Cedex 05, France.
}
\author{Leticia F. Cugliandolo}
\affiliation{$^1$Laboratoire de Physique Th\'eorique et Hautes Energies,
Universit\'e Pierre et Marie Curie -- Paris VI, 
4 Place Jussieu, 
75252 Paris Cedex 05, France.
}

\begin{abstract}
Impurity motion in one dimensional ultra cold quantum liquids confined in an optical trap has attracted much interest recently. 
As a step towards its full understanding, we construct a generating functional from which we 
derive the position {\it non equilibrium} correlation function of a quantum Brownian particle 
with general Gaussian non-factorizing initial conditions. We investigate the slow  
dynamics of a particle confined in a harmonic potential after a position measurement;
the rapid relaxation of a particle trapped in a harmonic potential after a 
quantum quench realized as a sudden change in the potential parameters; 
and the evolution of an impurity in contact with a one dimensional bosonic  quantum gas. We argue that such an impurity-Luttinger 
liquid system, which has been recently realized experimentally, admits a simple modeling as quantum Brownian motion in a 
super Ohmic bath. 
\end{abstract}

\maketitle

\newpage

\section{Introduction}\label{sec:intro}

Quantum Brownian motion has  been the starting point for the understanding of more complex dissipative quantum 
systems~\cite{We08}. Applications to quantum tunnel junctions~\cite{SchZaik90}, dissipative two--state systems~\cite{CalLegg87} 
and reaction--rate theory~\cite{Hanggi90} are just a few among many. In its simplest form, as proposed in the 
founding papers~\cite{FeynVern63,Schwinger61, CalLegg83}, the environment induced dissipation is modeled 
by an ensemble of quantum harmonic oscillators linearly coupled to the particle of interest. So far, in most  studies of the 
dissipative dynamics of a harmonically confined~\cite{Schwinger61,Agarwal71}  or a free~\cite{Grscin88,Pottier00} quantum particle, 
the quantity of interest has been the reduced density matrix that  
is obtained by tracing away the bath degrees of freedom in the  density matrix of the coupled system. 
For generic initial conditions this quantity has been obtained with the help of functional integral 
methods~\cite{Grscin88,Ingold02}. An alternative simpler, though in general only approximate, description of the reduced density 
matrix is given by a master equation. For factorizing initial conditions \cite{HuPaz92} and thermalized initial 
conditions \cite{Grab97} an exact master equation can be obtained. 
However, it is also known that there cannot be a master equation -- in the form of a partial differential equation local in time -- 
for arbitrary initial conditions~\cite{Grab97}.
The alternative quantum Langevin approach \cite{Ford65} extensively used in quantum optics \cite{Gardiner04} is not sufficiently 
powerful either, for quite 
the same reason: only a few special initial conditions can be successfully treated within this approach and the quantum noise statistics 
are not tractable in the generic case.

Non equilibrium correlation functions have been recently observed in cold atoms experiments on the dynamics of an \emph{impurity} atom 
moving in a one dimensional (1D) quantum liquid~\cite{Giamarchi11,Giamarchi07,Palzer09}. Both the impurity and the quantum liquid are confined in an optical harmonic trap so that the impurity motion resembles the dynamics of a damped quantum harmonic oscillator. In \cite{Johnson12} the authors attempted to describe the impurity dynamics within the Gross-Pitaevskii approach at zero temperature. In this paper we will follow an alternative way by applying quantum Brownian motion theory to the impurity problem. 
More specifically, we will generalize the path integral formalism in~\cite{Grscin88} to the language of \emph{generating functionals} 
commonly used in quantum field theory.  From the generating functional  we will easily deduce the \emph{non-equilibrium 
correlation functions} for generic non-factorizing Gaussian initial conditions (for a stochastic description of open quantum 
systems \emph{via} a generating functional, see \cite{Calzetta02}) that \emph{cannot} be obtained within the density matrix formalism. 
We will use this technique to treat three problems: the relaxation dynamics of a particle 
confined in a harmonic potential after a position measurement performed at the ``initial time"; the relaxation 
dynamics of a particle after an abrupt change in the parameters of the confining potential; the motion of an 
impurity~\cite{RoschKopp95,SchiroZwierlein09,Pita04,Caux09,Palzer09}
in contact with a 1D quantum gas all confined in a harmonic potential. In the latter case, we will explicitly 
compare our theoretical findings to recent experimental results \cite{Giamarchi11} on ultra cold quantum gases.
We will argue that, although the impurity-Luttinger liquid 
system is described by the Fr\"ohlich polaron Hamiltonian, many aspects of the impurity dynamics can be understood in the 
framework of quantum Brownian motion. 

More precisely, the paper is organized as follows. In Sec.~\ref{sec:model} we present the model and we review the main results 
obtained in~\cite{Grscin88}. In Sec.~\ref{genfunc} we employ path integral methods to derive the generating functional of 
out of equilibrium correlations. Our 
results cover both factorized and non-factorized Gaussian initial conditions as well as the effects of an initial 
position measurement performed on the particle. 
In Sec.~\ref{noneq} we study the equilibration processes after an initial position measurement and after a quench in the harmonic 
potential and we derive the equilibration times for low and high bath temperatures. In Sec.~\ref{luttinger} we apply our formalism 
to impurity motion in a $\rm 1D$ quantum gas described by the Luttinger theory. 
To keep the discussion simple, we choose to use a simplified modeling of the experiment in which we neglect
polaronic effects~\cite{Devreese09,Tempere09} as well as the possible renormalization of the external potential~\cite{Giamarchi11}.  
These subtle effects will be analyzed elsewhere. The Luttinger liquid is found to behave as 
an exotic quantum bath of harmonic oscillators with a highly \emph{non Ohmic spectral density} and non-linearly coupled to the particle. 
This is shown to lead to the curious behavior that 
the oscillator frequency can \emph{increase} upon increasing the coupling constant between the ``bath'' and the impurity in strong 
contrast to the behavior of an Ohmic damped oscillator. We further calculate the non equilibrium equal time correlation function (the 
variance of the position) and we compare our theoretical results to experimental evidence.
In the last section, Sec.~\ref{conclude}, we conclude and we present further possible applications of our work.

\section{The model}
\label{sec:model}

We study the evolution of a particle of mass $M$ evolving in a (possibly time-dependent) 
potential $V(\hat q;t)$ where $\hat q$ is the position operator. The Brownian motion  stems from its 
interaction with a quantum heat bath which is usually modeled by an infinite set of harmonic oscillators 
linearly coupled to the position 
operator $\hat q$. The full system is then described by the Hamiltonian $\hat{\mathcal H} = \hat{\mathcal H}_S + \hat{\mathcal H}_B+ 
\hat{\mathcal H}_{SB}$, with
\begin{align}
  \hat{\mathcal H}_S[\hat q,\hat p] &= \frac{\hat p^2}{2M} + V(\hat q;t) - H(t) \hat q
  \label{eq:Hs} 
   \; , \\
  \hat{\mathcal H}_B[\{\hat x_n,\hat p_n\}] &= \sum_{n=1}^\infty \frac{\hat p_n^2}{2m_n} + 
  \frac{m_n\omega_n^2}{2}\hat x_n^2 \label{eq:Hb}
   \; ,
  \\
  \hat{\mathcal H}_{SB}[\{\hat x_n\},\hat q] &= -\hat q \sum_{n=1}^\infty c_n\hat x_n + \hat q^2 \sum_{n=1}^\infty
  \frac{c_n^2}{2m_n\omega_n^2}
\label{eq:Hsb}
   \; .
\end{align}
$\hat p$ is the momentum operator of the particle. The last term in Eq.~(\ref{eq:Hs}) introduces a 
time-dependent source $H(t)$, a c-number, that couples linearly to the particle's position $\hat q$.  
$\hat x_n$ and $\hat p_n$ are the position and momentum operators
of the $n$--th harmonic oscillator, with mass and frequency $m_n$ and $\omega_n$, respectively.
$c_n$ is the coupling strength between the particle and the $n$-th oscillator's position. The last 
term in Eq.~(\ref{eq:Hsb}) compensates for the bath-induced renormalization of the potential. Indeed, the sum of Eqs.~(\ref{eq:Hb}) and~(\ref{eq:Hsb}) can be rewritten as
\begin{equation}
\hat{\mathcal H}_B+\hat{\mathcal H}_{SB} = \sum_{n=1}^\infty \frac{\hat p_n^2}{2m_n} + \frac{m_n\omega_n^2}{2} \ 
[\hat x_n - \frac{c_n}{m_n
\omega_n^2}\hat q]^2 \; ,
\end{equation}
which shows the absence of any drift force induced by the bath and ensures that $V(\hat q,t)$
corresponds to the physical potential right from the start.
The model Hamiltonian Eqs.~(\ref{eq:Hb})-(\ref{eq:Hsb}) has been widely used in the literature as a generic model 
for the dissipative dynamics of a quantum particle \cite{CalLegg83,Grscin88,We08}. 

In the Heisenberg representation the time evolution of all possible observables $\hat A$ is governed by
\begin{align}\label{eq:sch}
  \hat A(t) = &\left[\hat{\mathcal T} \exp\left( -\frac{i}{\hbar} \int_0^t dt' \ \hat{\mathcal H}(t')\right)\right]^\dagger 
  \hat A \nonumber\\
&\times \left[\hat{\mathcal T} \exp\left(-\frac{i}{\hbar}\int_0^t dt' \ \hat{\mathcal H}(t')\right)\right]
\; ,
\end{align}
with $\hat{\mathcal T}$ the time-ordering operator.
By introducing the density matrix of the initial state $\hat\rho_0$ 
the $N$-time average of a set $\{\hat A_i\}$ of $N$ operators,
is
\begin{align}\label{eq:av2}
  \langle \hat A_N(t_N) &\hat A_{N-1}(t_{N-1}) \cdots \hat A_1(t_1) \rangle = \nonumber\\
  &
  \qquad \mbox{Tr}\left[\hat A_N(t_N) \cdots \hat A_1(t_1) \hat\rho_0 \right]\; ,
\end{align}
where we took the product of the $\hat A_i$s to be time ordered (with $t_N \ge t_{N-1}\ge\ldots\ge t_1$) so that we 
can more easily make the connection between Eq.~(\ref{eq:av2}) and its path integral representation. We 
assumed that $\mbox{Tr}\hat\rho_0$ is normalized to one. Note that for a generic initial matrix $\hat\rho_0$ this,
as well as any other, correlation function is not necessarily stationary, i.e., it may depend on the $N$ times 
explicitly.

In all cases the model has to be supplemented by information on the initial condition 
of the coupled system. These are incorporated in the initial density matrix 
$\hat\rho_0$. \emph{Equilibrium dynamics} can be studied by 
choosing $\hat \rho_0$ to be the Boltzmann weight, that is 
\begin{equation}
\label{rho0thermal}
\hat \rho_0 = \exp(-\beta\hat{\mathcal H})\; ,
\end{equation}
where $\hat{\mathcal H}$ is the \emph{full coupled} Hamiltonian and the normalization constant has been ignored. 
This truly equilibrium density matrix has to be 
distinguished from $\hat \rho_0 = e^{-\beta \hat{\mathcal H}_S} \otimes  e^{-\beta \hat{\mathcal H}_B}$, a case in which
each component of the ``universe'' (the whole particle--bath system) is in equilibrium on its own at the 
same temperature. This subtle point is often overlooked  in the literature. 

\emph{Non equilibrium dynamics} can be studied whenever the initial density matrix is not of the form 
in Eq.~(\ref{rho0thermal}). 
 The simplest choice is  an initial product state for which the initial density matrix factorizes into two contributions 
 $\hat\rho_{S0}$ and $\hat\rho_{B0}$ which solely depend on particle and bath variables, respectively:
\begin{equation}\label{eq:initialfac}
  \hat\rho_0 = \hat\rho_{S0} \otimes \hat\rho_{B0}\; .
\end{equation}
Brownian motion~\cite{Hanggi05, We08}  
as well as the dynamics of more complex macroscopic 
systems~\cite{Cugliandolo06,Culo98,Culo99,Cugrlolosa,Kech,Kechye,Bipa,Arbicu10,Arbicu10b}
with a factorized initial density matrix have been studied in a variety of physical situations.
However, in many cases it is not appropriate to assume Eq.~(\ref{eq:initialfac}) since  one has
no command over the bath and it is impossible to ``switch it on and off'' at will. In addition, 
with recent developments in cold atom 
experiments, new classes of initial conditions become of relevance.

The first one covers all situations in which the particle is in equilibrium in a potential and either it is released or 
the potential is suddenly modified at $t = 0$. In this case Eq.~(\ref{rho0thermal}) holds with $\hat{\mathcal H}$ replaced by 
$\hat{\mathcal H}_0 \; (\neq  \hat{\mathcal H})$ describing the initial state.

The second class concerns all situations in which the position of the free particle is measured at $t=0$. This procedure
projects the initial density matrix onto the quantum states of the measurement 
outcome. We focus on the case where no quantum quench is performed in addition to the position measurement so that 
$\hat{\mathcal H}_0 = \hat{\mathcal H}$. If the position is exactly determined at $t=0$ 
the initial density matrix is
\begin{equation}
\label{eq:init3}
\hat \rho_0 = \hat\Pi(q) e^{-\beta \hat{\mathcal H}} \hat\Pi(q)\; 
\qquad \mbox{with} \qquad \hat\Pi(q) = |q\rangle\langle q|
\end{equation}
the projection operator onto the state $|q\rangle$. If, instead, we take the measured position of the particle to be Gaussian distributed around $q_m$ the projection operator takes the form
\begin{eqnarray}\label{eq:proj}
\hat\Pi(q_m) = \int{\rm d} q \ e^{-\frac{(q-q_m)^2}{4\Delta^2}} \ |q\rangle\langle q|\; ,
\end{eqnarray}
where $\Delta$ measures the uncertainty of the particle's position at $t=0$. Once again we neglected the irrelevant normalization factor. 

A third important class of initial conditions are the factorized density operators, see Eq.~(\ref{eq:initialfac}), in which the initial state 
of the system is a pure state. Since any state can be expanded in terms of displaced Gaussians (or \emph{coherent states}) it 
suffices to consider initial states of the form
\begin{eqnarray}\label{fac}
\hat\rho_{S0} = |\psi\rangle\langle\varphi|\; ,
\end{eqnarray}
where 
\begin{eqnarray}
\psi(q) = e^{-\frac{(q-q_m)^2}{4\Delta^2}}\;\;\;\mathrm{and}\;\;\;\varphi(q') = e^{-\frac{(q'-q'_m)^2}{4\Delta^2}}\; ,
\end{eqnarray}
to cover the whole class of initially factorized pure states.

In this article we derive a generating functional that allows us to obtain the $N$-time correlators for these types of initial conditions.
We are mainly interested in the evolution and averages of the \emph{particle's position observables} for which 
$\hat A = A[\hat q]$ with some function $A$ depending on the position $\hat q$ of the particle. Note that due to the coupling of the 
bath to the particle's position the momentum dynamics follow from the Heisenberg equation $M \partial_t \hat q = \hat p$. 
Therefore, by focusing on the particle position operator we simultaneously describe the dynamics of the particle's momentum. 
While in \cite{Grscin88} 
the authors derived an explicit expression for the equilibrium correlation functions the generating functional will allow us to go beyond the equilibrium 
case.

\section{The generating functional}\label{genfunc}

In this section we derive the generating functional of all (non-equilibrium) correlation functions. 
This goes beyond the analysis in \cite{Grscin88} where explicit expressions for equilibrium correlation functions were given. 
More precisely, we derive a functional of two time-dependent sources ${\cal J}[F,G]$ such that the two-time correlation is given by 
\begin{eqnarray}
&& \langle \hat q(t) \hat q(t') \rangle =
\nonumber\\
&& 
\qquad 
\left.
\frac{\hbar}{i}\frac{\delta}{\delta G(t)}\frac{\hbar}{i}
\left[\frac{\delta}{\delta G(t')}+\frac{\delta}{2\delta F(t')}\right] e^{\mathcal J[F,G]} \right|_{F,G\equiv 0}
\end{eqnarray}
 and similarly for higher order correlations. 

We obtain the path integral formulation of the generating functional by making use of the coherent states of the bath 
variables $|\{\xi_{n,f}\}\rangle$ which 
are defined in  App.~\ref{app1}. The ensuing functional integration includes paths over particle and bath variables. 
Since we are not interested in the degrees of freedom of the bath, we average over all bath variables to find a 
``reduced action''  that only depends on the particle variables. In the special cases discussed below (e.g., harmonic
potential) the remaining path-integrals can also be performed and the functional ${\cal J}$ can be fully determined. 
In this section we sketch all steps in the derivation. Further technical details are reported in App.~\ref{app1}.
The reader who is not interested in these technical details 
can jump directly to Eq.~(\ref{eq:Jfinal}) where its rather lengthy final expression is given.

\subsection{The density matrix}\label{sec:reddens}

In terms of the product states $|q,\{\xi_{n,f}\}\rangle$ between the particle and the bath eigenstates the 
matrix elements of the time evolution operator read
\begin{align}
  \mathcal K(q_f,\{\xi_{n,f}\};&q_i,\{\xi_{n,i}\};t) \equiv \nonumber\\
&\langle q_f,\{\xi_{n,f}\}|
  \hat{\mathcal T} e^{-\frac{i}{\hbar} \int_0^t dt' \ \hat{\mathcal H}(t')}
  |q_i,\{\xi_{n,i}\}
  \rangle
  \label{eq:K}
\end{align}
and of its Hermitian conjugate
\begin{eqnarray}
&&
  \mathcal K^*(q'_f,\{\xi'_{n,f}\}; q'_i,\{\xi'_{n,i}\};t) \equiv 
  \qquad\qquad\qquad\qquad
  \nonumber\\
&&
\qquad\qquad
\langle q'_i,\{\xi'_{n,i}\}|\hat{\mathcal T}^\dagger 
  e^{\frac{i}{\hbar}  \int_0^t dt' \ \hat{\mathcal H}(t')} 
  |q'_f,\{\xi'_{n,f}\}\rangle \; .
  \label{eq:Kstar}
\end{eqnarray}   
$\hat{ \mathcal T}^\dagger$ is the anti-chronological time ordering operator.
The elements of the time-dependent density matrix, $\hat \rho(t) \equiv \mathcal K \hat\rho_0 \mathcal K^*$, 
are given by 
\begin{align}
&
\mathcal W(q_f,\{\xi_{n,f}\};q_f',\{\xi'_{n,f}\};t) \equiv 
\langle q_f,\{\xi_{n,f}\}|\ \hat\rho(t)\ |q'_f,\{\xi'_{n,f}\}\rangle
\nonumber\\ 
&
\qquad
= 
\int{\rm d} q_i{\rm d} q'_i{\rm d}\xi_i{\rm d}\xi_i' \ \mathcal K(q_f,\{\xi_{n,f}\};q_i,\{\xi_{n,i}\};t)
\nonumber\\
& 
\qquad 
\times
\mathcal W(q_i,\{\xi_{n,i}\};q'_i,\{\xi'_{n,i}\})
\
\mathcal K^*(q'_f,\{\xi'_{n,f}\};q'_i,\{\xi'_{n,i}\};t) 
\nonumber\\
&
\label{eq:densityT}
\; ,
\end{align}
where  the matrix elements of the initial density matrix have been denoted by 
\begin{eqnarray}
\label{eq:W113}
\mathcal W(q_i,\{\xi_{n,i}\};q_i',\{\xi_{n,i}'\}) = \langle q_i,\{\xi_{n,i}\}|\ \hat\rho_0 \ |q_i',\{\xi_{n,i}'\}\rangle
\;\;\;\;\;
\end{eqnarray}
and we  used the short-hand notation
\begin{eqnarray}\label{eq:dxi}
{\rm d}\xi_i = \prod_{n=1}^\infty\ e^{-\xi_{n,i}^*\xi_{n,i}}{\rm d}\xi_{n,i}^*{\rm d}\xi_{n,i} 
\; .
\end{eqnarray}
The path integral representations of 
$\mathcal K$ and $\mathcal K^*$ are 
\begin{eqnarray*}
\label{eq:K-path}
&&
{\mathcal K}(q_f,\{\xi_{n,f}\};q_i,\{\xi_{n,i}\};t) = \! \int \! \mathcal D q^+\mathcal D \xi^+ e^{\frac{i}{\hbar}\mathcal S[q^+,\{\xi^+_n\}]} \; , \\
\label{eq:Kstar-path}
&&
{\mathcal K}^*(q'_f,\{\xi'_{n,f}\};q'_i,\{\xi'_{n,i}\};t) = \! \int \! \mathcal D q^- \mathcal D \xi^- e^{-\frac{i}{\hbar}\mathcal S^*[q^-,\{\xi^-_n\}]} \; ,
\nonumber
\end{eqnarray*} 
where we made clear with the superscripts $^+$ and $^-$ which paths belong to $\mathcal K$ and $\mathcal K^*$, respectively. 
The functional integration measures are defined in App.~\ref{app1}.

\subsection{Reduced density matrix for a system initially coupled to an equilibrium bath}
\label{roleinit}

The time-dependent density matrix in Eq.~(\ref{eq:densityT}) still contains information about the degrees 
of freedom of the bath which we are not interested in. Therefore, we average (trace) over all the bath variables to find a reduced density 
matrix  
%
%
that depends only on the particle variables and the external sources.

We are interested in a system that is initially coupled to an equilibrium bath. Therefore, 
\begin{equation}
 \hat{\mathcal H}_{B0} = \hat{\mathcal H}_B
  \; , 
  \qquad\qquad
  \hat{\mathcal H}_{SB0} = \hat{\mathcal H}_{SB} 
  \; ,
\end{equation}
where all initial Hamiltonians are labeled with a subscript $_0$. At this point it is not necessary to make  $\hat{\mathcal H}_{S0}$ explicit since this term 
involves only particle variables that are not affected by the trace over the bath variables.
The matrix element of the initial density operator Eq.~(\ref{eq:W113}) in Eq.~(\ref{eq:densityT}) can be represented by an 
imaginary time path integral
\begin{eqnarray*}
\mathcal W(q_i,\{\xi_{n,i}\};q'_i,\{\xi'_{n,i}\}) = 
\int\mathcal D q^0 \mathcal D \xi^0 \ 
  e^{-\frac{1}{\hbar}\mathcal S_0[q^0,\{\xi^0_n\}]}
  \; ,
\end{eqnarray*}
where the initial action $\mathcal S_0$ is in general different from $\mathcal S$ reflecting the fact that 
$\hat{\mathcal H}_0 \neq \hat{\mathcal H}$. The reduced density matrix can now be recast as 
\begin{align*}
&\mathcal W(q_f;q_f';t) \equiv \int{\rm d} q_i{\rm d} q'_i \mathcal D q^0 \mathcal D q^+ \mathcal D q^-
\nonumber\\
&\qquad\qquad\times e^{\frac{i}{\hbar}\mathcal S_S[q^+]-\frac{i}{\hbar}\mathcal S_S[q^-] 
- \frac{1}{\hbar} {\mathcal S_{S}}_0[q^0]} \ \mathcal F[q^+,q^-,q^0] \; , \nonumber
\end{align*}
where $\mathcal F[q^+,q^-,q^0]$ is the  ``influence functional'' that  depends only on the particle variables, 
as also do  $\mathcal S_S$ and ${\mathcal S_{S}}_0$. The path integral runs over all paths with 
\begin{align*}
&q^+(t) = q_f\;  , \;\;\;q^+(0) = q_i\;  , \;\;\;q^-(t) = q_f'\;  , \\
&q^-(0) = q_i'\; , \;\;\;q^0(\beta\hbar) = q^-(0)\; , \;\;\;q^0(0) = q^+(0)\;  ,
\end{align*}
which is the reason for the name ``closed-time path integral''. 
It is convenient to introduce the linear combinations
\begin{eqnarray}\label{xxbar}
x = (q^++q^-)/2 \qquad \mathrm{and} \qquad  \bar x = q^+ - q^-\; .
\end{eqnarray}
The calculation of $\mathcal F$ can be found in App.~\ref{app2} or in \cite{Grscin88}; the result reads 
\begin{eqnarray}
\mathcal F[x,\bar x,q^0] = e^{\frac{i}{\hbar} \Phi[x,\bar x,q^0]}
 \; ,
\end{eqnarray}
with
\begin{eqnarray}
\label{eq:phi2}
&&\Phi[x,\bar x,q^0] =  \frac{i}{2} \int_0^{\beta\hbar}{\rm d}\tau{\rm d}\sigma \ k(\tau-\sigma)q^0(\tau)q^0(\sigma) 
\nonumber\\
&&
\;\;\;\;\; + \int_0^{\beta\hbar}{\rm d}\tau\int_0^t{\rm d} s \ K^*(s-i\tau)q^0(\tau)\bar x(s) 
\nonumber\\
&&
\;\;\;\;\; - \frac{i}{2}\int_0^t{\rm d} s{\rm d} u \ K_R(s-u)\bar x(s) \bar x(u) 
\nonumber\\
&& 
\;\;\;\;\; - M \int_0^t{\rm d} s \ \bar x(s)\frac{{\rm d}}{{\rm d} s}\int_0^s{\rm d} u \ \gamma(s-u) x(u)
\; .
\end{eqnarray}
The kernels $K(\theta)$, $\gamma(t)$ and $k(t)$ are defined in Eqs.~(\ref{eq:kernel}),~(\ref{eq:gamma}) and~(\ref{eq:smallk}), respectively. 
$K_R$ denotes the real part of $K$. Note that $\Phi[x,\bar x,q^0]$ depends on the fixed 
``end-points" $t$ and $\beta\hbar$ of the closed-time path.

Expected values evaluated at different times are now expressed in terms of a path-integral
over $q^0$, $x$ and $\bar x$ with an effective action $\Sigma$,  
\begin{align}
\label{eq:corr1Sigma}
&\langle  \dots \rangle = \int{\rm d} x_f{\rm d} x_i{\rm d} \bar x_i\ \int' \mathcal D x\mathcal D \bar x \mathcal D q^0 
\ \dots \ e^{\frac{i}{\hbar}\Sigma[x,\bar x,q^0]} \; ,
\end{align}
where $\Sigma[x,\bar x,q^0] $ is given by
\begin{align}\label{eq:Sigma}
&\Sigma[x,\bar x,q^0] = \nonumber\\
&\Phi[x,\bar x,q^0] + i {\mathcal S_S}_0[q^0] + \mathcal S_S[x + \bar x/2] \nonumber
- \mathcal S_S[x - \bar x/2] 
\nonumber\\
&= \Phi[x,\bar x,q^0] + i\int_0^{\beta\hbar}{\rm d}\tau \left[ \frac{M_0}{2} (\dot{q}^0)^2 + V_0(q^0)\right] 
\nonumber\\
&+ \int_0^t{\rm d} s \left[ M \dot{\bar x} \dot x - V_H\left(x+\frac{\bar x}{2};s\right) + V_{H'}\left(x - \frac{\bar x}{2};s\right)\right] \; .
\end{align}
We introduced the initial mass $M_0$ of the particle and the initial potential $V_0$ that are 
in general different from the ``bulk'' mass $M$ and potential $V$. This allows for 
quenches in these parameters. Note that the case in which the initial state is a pure state [e.g. Eq.~(\ref{fac})]
can be easily recovered by setting $M_0 = 0$ and $V_0 = 0$ or, equivalently, by noting that the path 
$q^0$ shrinks identically to zero (since there is no initial Hamiltonian for this simple type of initial condition).

The superscripts in the path integral in Eq.~(\ref{eq:corr1Sigma})  
remind us of the constraint that the paths are subject to. One has
\begin{eqnarray}
&& x(0) = x_i , \;\; \bar x(0)=\bar x_i , 
  \label{eq:constraints1}
\nonumber\\
&& x(t) = x_f  , \;\; q^0(0) = q^+(0) = x_i + \frac{\bar x_i}{2} ,   
\label{eq:constraints2}
\\
&& \bar x(t) = 0 , \;\; q^0(\beta\hbar) = q^-(0) = x_i - \frac{\bar x_i}{2} 
\; .
\nonumber
\end{eqnarray}  
Note that due to the periodic boundary conditions of the trace $\bar x_f = q^+(t) - q^-(t) = 0$. 

\subsection{Generic Gaussian initial conditions}

It is very easy to include the change of $\Phi[x,\bar x,q^0]$ induced by the initial position measurement in Eq.~(\ref{eq:init3}). 
By using the 
explicit Gaussian form of the projector $\hat\Pi(q_m)$ [see Eq.~(\ref{eq:proj})] the dependence on the initial measurement can be 
simply incorporated in 
$\Sigma[x,\bar x,q^0]$ by an additional term of the form
\begin{eqnarray}
\label{eq:measurement}
&& 
\frac{i\hbar}{4\Delta^2} 
\!
\left[ 
(x_i + \frac{\bar x_i}{2}-q_m)^2 + (x_i - \frac{\bar x_i}{2} - q_m)^2 \right] 
\nonumber\\
&& \;\;\;\; \qquad
= 
\frac{i\hbar}{2\Delta^2} \! \left[(x_i - q_m)^2 + \frac{\bar x_i ^2}{4}\right] 
\!
\; .
\end{eqnarray}   
In the limit of strong uncertainty $\Delta \to \infty$ the effect of the initial measurement is blurred.

In order to recover the case where the initial state of the system is pure and decouples from the environment [which 
corresponds to the factorized initial density matrix with $\hat\rho_{S0}$ given by Eq.~(\ref{fac})] 
the action in Eq.~(\ref{eq:Sigma}) has to be supplemented by 
\begin{align}
\label{initialpure}
& 
\frac{i\hbar}{4\Delta^2}
\left[
(x_i + \frac{\bar x_i}{2}-q_m)^2 + (x_i - \frac{\bar x_i}{2} - q'_m)^2 
\right]
\nonumber\\
&
= \frac{i\hbar}{2\Delta^2}\left[x_i^2+\frac{\bar x_i^2}{4} +x_m^2 + \frac{{\bar x}_m^2}{4} - 2x_mx_i  -\frac{1}{2} \bar x_m \bar x_i \right]
 \; ,
 \nonumber\\
\end{align}
with the  notation
\begin{eqnarray}
x_m = (q_m+q'_m)/2\;\;\;\mathrm{and}\;\;\;\bar x_m = (q_m-q'_m)  \; .
\end{eqnarray}
Since Eq.~(\ref{eq:measurement}) is a special case of Eq.~(\ref{initialpure}) 
corresponding to $q_m=q_m'$ (or $x_m=q_m$ and $\bar x_m=0$)
we will work with the latter in the following. The relevant cases can then be selected by taking 
simple limits.

In  the following expressions we will write only the terms that depend on $x_i$ or $\bar x_i$ since 
the ones depending on $x_m$ and $\bar x_m$ contribute only to an overall constant.
 
\subsection{The sources}

The source term appears as $\int dt' H(t') q^+(t')$ in ${\cal K}$ [see Eq.~(\ref{eq:K})]  
and as $-\int dt' H(t') q^-(t')$ in ${\cal K}^*$ [see Eq.~(\ref{eq:Kstar})]. For convenience, we 
distinguished the function existing on the positive running branch of the closed time contour, 
which we still call $H(t)$, from the one existing on the negative running branch of the same
contour, which we  call $H'(t)$. This implies that the potentials in Eq.~(\ref{eq:Sigma}) are 
given by $V_H(y) = V(y) - H y$ and $V_{H'}(y) = V(y) - H' y$.

After the transformation of variables in Eq.~(\ref{xxbar})
we obtain two external time-dependent sources $F(s)=[H(s)+H'(s)]/2$ and 
$G(s)=[H(s)-H'(s)]$ which couple linearly to the variables 
$\bar x(s)$ and $x(s)$, respectively. All correlation functions can be computed 
from the generating functional $\mathcal J[F,G]$ as derivatives of $\mathcal J$ with 
respect to $F$ or $G$ evaluated at $F=G=0$. A physical force is represented by $H(s)=H'(s)$, that is by $F(s) \neq 0$ and $G(s) = 0$. 
Therefore, the linear response of the mean value Eq.~(\ref{eq:corr1Sigma}) to an external force can be 
obtained for $F(s) \neq 0$. 

The generating functional, that is to say, 
the trace over the reduced density matrix in the presence of the external sources  reads
\begin{eqnarray*}
&
e^{\mathcal J[F,G]} \sim \int{\rm d} x_i{\rm d}\bar x_i{\rm d} x_f \int'\mathcal D x\mathcal D \bar x \mathcal D q^0
\ e^{\frac{i}{\hbar}\Sigma[x,\bar x, q^0,x_i,x_f,\bar x_i]} \; ,
\end{eqnarray*}
where the path integral is subject to the constraints in Eqs.~(\ref{eq:constraints2}). 
The overall normalization factor depends on $t$, $\beta$ and all parameters in the model but not on the fields. We can now write
\begin{align}\label{eq:rf}
&\langle \hat q(t) \rangle = \frac{\hbar}{i}\frac{\delta}{\delta G(t)} \left. e^{\mathcal J[F,G]}\right|_{F\equiv 0,G\equiv 0}
\end{align}
and
\begin{align}\label{eq:cf}
&\langle \hat q(t)\hat q(t') \rangle =\nonumber \\
&
\qquad
\frac{\hbar}{i}\frac{\delta}{\delta G(t)}\frac{\hbar}{i}\left[\frac{\delta}{\delta G(t')}+\frac{\delta}{2\delta F(t')}\right]
\left. e^{\mathcal J[F,G]}\right|_{F,G\equiv 0}
\end{align}
and all other correlation functions can be obtained in a similar way by noting that $q^+(t) = q_f = x_f$ and $q^+(t') = x(t) + \bar x(t)/2$. At this point it has 
become obvious why two sources are needed in order to obtain all non-equilibrium correlation functions. 

\subsection{The harmonic case}\label{sec33}

To go further 
we restrict ourselves to the study of a quantum Brownian particle in a harmonic potential for which
\begin{eqnarray}\label{eq:pot1}
- V(x+\bar x/2;s) + V(x - \bar x/2;s) = -M\omega^2x\bar x
\end{eqnarray}
and
\begin{eqnarray}\label{eq:pot2}
V_0(q^0) = \frac{1}{2} M_0\omega_0^2(q^0)^2 \; .
\end{eqnarray}
The choice of a quadratic potential renders the problem analytically solvable. The generating functional
can be calculated by simply evaluating the action on its minimal action path (over the initial condition
branch and the time-dependent branches) as Gaussian fluctuations yield only pre-factors that are independent
of the sources and can be determined at the end of the calculation from the normalization of the density matrix.
Note that, although both initial and bulk potentials are harmonic, they are not necessarily the same thus
allowing for the study of quantum quenches. 

\subsection{Integration over the initial condition}\label{sec:intinitial}

We first treat the contribution of the initial condition path $q^0$ in Eq.~(\ref{eq:Sigma}).
The equation of motion for $q^0$ can be easily obtained from Eq.~(\ref{eq:Sigma}):
\begin{align}
\nonumber
M_0\ddot{q}^0(\tau) &- \int_0^{\beta\hbar} \!\!\! {\rm d}\sigma \ k(\tau-\sigma) q^0(\sigma) - M_0\omega_0^2 q^0(\tau) \\
&= -i\int_0^t \! {\rm d} s \ K^*(s-i\tau)\bar x(s) 
\; ,
\end{align}
with the fixed end-points $q^0(0) = q^+_i$ and $q^0(\beta\hbar) = q^-_i$. As the $q^0$ path is part of the whole closed-time path it 
implicitly depends on the fixed end-time $t$ as well. 
In \cite{Grscin88} one can find a detailed analysis of this equation of motion
which uses a Fourier expansion of the path 
$q^0(\tau)$ on the interval $[0,\beta\hbar]$. 
By using the results found therein we obtain
\begin{widetext}
\begin{eqnarray}
\label{eq:Sigma2}
\Sigma[x,\bar x,x_i,x_f,\bar x_i] 
&=& 
\frac{i}{2M_0}\int_0^t{\rm d} s{\rm d} u \ R'(s,u)\bar x(s) \bar x(u) 
+ iM_0\left[\frac{1}{2\Lambda}x_i^2 + \frac{\Omega}{2}\bar x_i^2\right] 
+ \frac{i\hbar}{2\Delta^2}\left[x_i^2+\frac{\bar x_i^2}{4} - 2x_mx_i - \frac{1}{2}\bar x_m\bar x_i \right] 
\nonumber\\
&&
+
\int_0^t{\rm d} s \ M\left[\dot{\bar x}\dot x - \omega^2\bar x x + 
\frac{1}{M}\bar F(s) \bar x(s) + \frac{1}{M}G(s)x(s) 
- \bar x(s) \frac{{\rm d}}{{\rm d} s}\int_0^s{\rm d} u \gamma(s-u)x(u)\right] 
\; .
\end{eqnarray}
\end{widetext}
We introduced the complex ``force''
\begin{eqnarray}\label{eq:force}
\bar F(s) = F(s) + x_iC_1(s) - i\bar x_i C_2(s)
\; , 
\end{eqnarray} 
with the functions $C_1$ and $C_2$
\begin{align}\label{C1C2}
  C_1(s) &= \frac{1}{\beta\hbar\Lambda} \sum_{k=-\infty}^\infty u_k g_k(s)
  \;, \nonumber\\
  C_2(s) &= \frac{1}{\beta\hbar} \sum_{k=-\infty}^\infty u_k\nu_k f_k(s)
  \; .
\end{align}
The constants $\Lambda$ and $\Omega$ are  given by
\begin{align}\label{eq:def1}
&\Lambda = \frac{1}{\beta\hbar} \sum_{k=-\infty}^\infty u_k \;\;\;\mathrm{and}\;\;\;\Omega = \frac{1}{\beta\hbar} \sum_{k=-\infty}^\infty u_k(\omega_0^2 + \zeta_k) 
\; ,
\end{align}
with
$u_k = (\omega_0^2 + \nu_k^2 + \zeta_k)^{-1}$,  
$\nu_k = 2\pi k/\beta\hbar $,
$\zeta_k = [M\gamma(0) - g_k(0)]/M_0$ [for the definition of $\gamma(t)$ see Eq.~(\ref{gamma}) below]
and 
\begin{eqnarray}
\label{eq:gk}
g_k(s) = \int_0^\infty\frac{{\rm d}\omega}{\pi} S(\omega)\frac{2\omega}{\omega^2 + \nu_k^2}\cos(\omega s)
\; , 
\\
\label{eq:fk}
f_k(s) = \int_0^\infty\frac{{\rm d}\omega}{\pi} S(\omega)\frac{2\nu_k}{\omega^2 + \nu_k^2}\sin(\omega s)\; ,
\end{eqnarray}
where $S(\omega)$ is the spectral density of the bath.
The two-time function $R'$ reads
\begin{align}\label{Rtilde}
R'(s,u) &= R(s,u) + M_0 K_R(s-u) 
\; , 
\nonumber\\
R(s,u) &= -\Lambda C_1(s)C_1(u) \nonumber\\
&+ \frac{1}{\beta\hbar}\sum_{k=-\infty}^\infty u_k\left[g_k(s)g_k(u)-f_k(s)f_k(u)\right] 
\; , 
\end{align}
with 
\begin{eqnarray}
K_R(s-u) = \frac{1}{\beta\hbar}\sum_k \ g_k(s-u) 
\end{eqnarray}
the real part of the kernel $K$. 
The time-dependent bath kernel $\gamma(s)$ is given by [see Eq.~(\ref{eq:gamma})]
\begin{eqnarray}
\label{gamma}
\gamma(s) 
&=& 
\frac{2}{M}\int_0^\infty\frac{{\rm d}\omega}{\pi} \frac{S(\omega)}{\omega} \cos(\omega s) \; .
\end{eqnarray}
The functions $C_1$ and $C_2$ as well as the kernel $R(s,u)$ are not to be confused with the correlation functions and the 
linear response function that will be denoted by $\mathcal C$ and $\mathcal R$, respectively.

\subsection{Real time minimal action paths with external sources}
\label{sec:realtime}

The equations of motion for $x(s)$ and $\bar x(s)$ read
\begin{align}\label{eq:motion2} 
  \ddot{x}(s) &+ \frac{{\rm d}}{{\rm d} s}\int_0^s{\rm d} u \ \gamma(s-u) x(u) + \omega^2 x(s)  
   \nonumber\\
  &=  \frac{\bar F(s)}{M} + \frac{i}{M M_0}\int_0^t{\rm d} u \ R'(s,u)\bar x(u)
  \; ,
\\
\label{eq:motion3}
 \ddot{\bar x}(s) &- \frac{{\rm d}}{{\rm d} s}\int_s^t{\rm d} u \ \gamma(u-s) \bar x(u) + \omega^2 \bar x(s) = \frac{G(s)}{M}
   .
\end{align}
The action $\Sigma$ evaluated along the minimal action paths can  be determined by 
inserting the solutions to Eqs.~(\ref{eq:motion2}) and~(\ref{eq:motion3}) into Eq.~(\ref{eq:Sigma2}). 
However, the authors of \cite{Grscin88} noted a 
simplification of the calculation which can be generalized to our case where the source $G(s)$ is also present 
(in \cite{Grscin88} no external source for $\bar x$ was used). 
The idea is the following.
After a partial integration in the second line of Eq.~(\ref{eq:Sigma2}) the action $\Sigma$ takes the form
\begin{eqnarray}
 &&\Sigma[x,\bar x,x_i,x_f,\bar x_i] =
\nonumber \\
 &&
 \qquad 
 -M \bar{x}_i \dot{x}_i - \frac{i}{2M_0}\int_0^t{\rm d} s{\rm d} u 
\ R'(s,u)\bar x(s) \bar x(u) 
 \nonumber
  \\
&&
\qquad
+ \int_0^t{\rm d} s\ G(s)x(s) + \mathrm{border}(x_i,x_f,\bar x_i)
\label{eq:102}
\end{eqnarray}
when evaluated along the minimal action paths determined by Eqs.~(\ref{eq:motion2}) and (\ref{eq:motion3}), 
where we used the boundary condition $\bar{x}_f = 0$. Here, 
$\mathrm{border}(x_i,x_f,\bar x_i)$ stands for all border terms in Eq.~(\ref{eq:Sigma2}). 

On the other hand, 
one can split the force Eq.~(\ref{eq:force}) into its real and imaginary parts $\bar F(s) = \bar{F}_R(s) + i\bar{F}_I(s)$. Then, the minimal 
action path $x(s)$ splits into $x(s) = x_R(s) + ix_I(s)$, where $x_I(s)$ satisfies the boundary conditions $x_I(0)=x_I(t)=0$. The trick is 
to show now that one can simply focus on the real part $x_R(s)$ of the minimal action path 
in order to obtain the complete stationary phase action. Indeed, if 
we evaluate the action Eq.~(\ref{eq:Sigma2}) only along the minimal $x_R(s)$ and $\bar x(s)$ 
we obtain 
\begin{eqnarray}\label{eq:103}
&&
\Sigma[x_R,\bar x, x_i, x_f, \overline x_i] 
= -M \bar{x}_i \dot{x}_{R,i} 
 + \int_0^t  {\rm d} s \ G(s)x_R(s) 
 \nonumber\\
 && 
 \qquad\qquad
+ \int_0^t  {\rm d} s\ \bar x(s) 
\left[\bar{F}_I(s) + \frac{i}{2M_0}\int_0^t  {\rm d} u \ R'(s,u)\bar x(u)\right] 
\nonumber\\
&&
\qquad\qquad
 + \mathrm{border}(x_i,x_f,\bar x_i)
\; ,
\end{eqnarray}
where we used the fact that $x_R(s)$ satisfies the real part of Eq.~(\ref{eq:motion2}). We now want to show that Eqs.~(\ref{eq:103}) and~(\ref{eq:102}) are indeed equal. With the help of the imaginary part of Eq.~(\ref{eq:motion2}) and the 
equation of motion~(\ref{eq:motion3}) we can easily prove by integration by parts that
\begin{align*}
\int_0^t \! {\rm d} s \bar x(s) \! &
\left[\bar{F}_I(s) + \frac{1}{M_0}\int_0^t \! {\rm d} u \ R'(s,u)\bar x(u)\right] 
\nonumber\\  
&= -M \bar{x}_i \dot{x}_{I,i} + \int_0^t \! {\rm d} s \ G(s)x_I(s)  
\; , 
\nonumber
\end{align*}
and by using this identity in Eq.~(\ref{eq:103}) we recover Eq.~(\ref{eq:102}). Therefore, the right-hand-side (rhs) of 
Eq.~(\ref{eq:102}) and the rhs of Eq.~(\ref{eq:103}) 
coincide. It is sufficient to evaluate 
the action Eq.~(\ref{eq:Sigma2}) along the real component $x_R(s)$ that satisfies a much simpler equation than $x(s)$. 

In terms of the end points $x_i$, $x_f$ and $\bar x_i$, the solutions to the real parts of Eqs.~(\ref{eq:motion2}) and~(\ref{eq:motion3}) read 
\begin{eqnarray}\label{solx}
x_R(s) &=& 
\frac{\mathcal G_+(s)}{\mathcal G_+(t)}x_f + 
\left[\dot{\mathcal G}_+(s) - \frac{\mathcal G_+(s)}{\mathcal G_+(t)}\dot{\mathcal G}_+(t)\right]x_i \nonumber\\
&& \;\;\; + \frac{1}{M}\int_0^s{\rm d} u\ \mathcal G_+(s-u) \bar{F}_R(u)
\nonumber\\
&& 
 \;\;\; 
 - \frac{1}{M} \frac{\mathcal G_+(s)}{\mathcal G_+(t)}\int_0^t{\rm d} u\ \mathcal G_+(t-u) \bar{F}_R(u)
\end{eqnarray}
and
\begin{eqnarray}
\label{eq:solxbar}
\bar x(s) &=& \frac{\mathcal G_+(t-s)}{\mathcal G_+(t)}\bar x_i + \frac{1}{M}\int_s^t{\rm d} u\ \mathcal G_+(u-s) G(u)
\nonumber\\
&&
 - 
 \frac{1}{M} \frac{\mathcal G_+(t-s)}{\mathcal G_+(t)}\int_0^t{\rm d} u\ \mathcal G_+(u) G(u)\; ,
\end{eqnarray}
where $\mathcal G_+(t)$ is a propagator that in Laplace-transform reads
\begin{eqnarray}\label{eq:Gplus}
\tilde{\mathcal G}_+(\lambda) = \frac{1}{\lambda^2 + \lambda\tilde\gamma(\lambda) + \omega^2} 
\; . 
\end{eqnarray}
From Eq.~(\ref{solx}) we immediately find, by using the boundary conditions  $\mathcal G_+(0) = 0$, $\dot{\mathcal G}_+(0) = 1$ 
and $\ddot{\mathcal G}_+(0) = 0$,
\begin{align}\label{eq:dotxi}
\dot{x}_{R,i} =
&
\frac{1}{\mathcal G_+(t)}x_{f} - \frac{\dot{\mathcal G}_+(t)}{\mathcal G_+(t)}x_i \nonumber\\
&-\frac{1}{\mathcal G_+(t)}\frac{1}{M}\int_0^t{\rm d} u \ \mathcal G_+(t-u)\bar{F}_R(u) \; .
\end{align}
Inserting the solutions to Eqs.~(\ref{solx}) and (\ref{eq:solxbar}),  
and Eq.~(\ref{eq:dotxi}) into Eq.~(\ref{eq:103}) we find an effective action $\Sigma[x_i,x_f,\bar x_i]$ that depends only on the end-points, $x_i$, $x_f$ and $\bar x_i$, and the external sources $F$ and $G$:
%
%
\begin{widetext}
\begin{eqnarray}
\label{eq:Sigma3}
\Sigma[x_i,x_f,\bar x_i] 
&= & 
- \frac{iM_0}{\epsilon^2}x_i x_m - \frac{iM_0}{4\epsilon^2}\bar x_i \bar x_m -M \bar x_i x_f \frac{1}{\mathcal G_+(t)} 
+ M\bar x_i x_i \frac{\dot{\mathcal G}_+(t)}{\mathcal G_+(t)} 
+ \frac{\bar x_i}{\mathcal G_+(t)}\int_0^t \!\! {\rm d} u \ \mathcal G_+(t-u)\bar F_R(u)  \nonumber\\
&&
- i\bar x_i \int_0^t \!\! {\rm d} s \ C_2(s)\bar x(s) 
+ \frac{iM_0}{2} \left(\frac{x_i^2}{\Lambda'} + \Omega' \bar x_i^2\right) 
+ \frac{i}{2M_0} \int_0^t{\rm d} s{\rm d} u \ R'(s,u) \bar x(s) \bar x(u) 
+ x_f \int_0^t{\rm d} s \frac{\mathcal G_+(s)}{\mathcal G_+(t)} G(s) \nonumber\\
&&
+ x_i \int_0^t{\rm d} s \ G(s)\left[\dot{\mathcal G}_+(s) - \frac{\mathcal G_+(s)}{\mathcal G_+(t)}\dot{\mathcal G}_+(t)\right] 
+ \frac{1}{M}\int_0^t{\rm d} s\int_0^s{\rm d} u \ \mathcal G_+(s-u) G(s) \bar F_R(u) \nonumber\\
&&
- \frac{1}{M}\int_0^t{\rm d} s{\rm d} u \frac{\mathcal G_+(t-u)}{\mathcal G_+(t)}\mathcal G_+(s) G(s) \bar F_R(u) 
 \; ,
\end{eqnarray}
\end{widetext}
with
\begin{eqnarray}
&& 
\frac{1}{\Lambda'} \equiv \frac{1}{\Lambda} + \frac{1}{\epsilon^2}, 
\qquad \Omega' \equiv
\Omega + \frac{1}{4\epsilon^2}\; ,
\nonumber\\
&& 
\mathrm{and}\qquad\qquad \epsilon^2 \equiv \frac{M_0\Delta^2}{\hbar} 
 \; .
\label{eq:definitions}
\end{eqnarray}
For the sake of a clear presentation we have not replaced $\bar x(s)$ and $\bar F_R$ by their 
corresponding expressions in terms of the end-points yet. 

\subsection{Integration over the end-points}

In order to find the final expression for $\mathcal J[F,G]$ 
 we still have to integrate over the end-points $x_i$, $x_f$ and $\bar x_i$. Since 
 the exponent $\Sigma[x_i,x_f,\bar x_i]$ is linear in $x_f$ the integration over 
 this variable generates a $\delta$--function of the form
\begin{eqnarray}
\delta\left[\bar x_i - \frac{1}{M}\int_0^t{\rm d} s \ \mathcal G_+(s)G(s)\right]
 \; , 
\end{eqnarray}
up to a factor not depending on the end-points
which, in combination with the integration over $\bar x_i$, enforces a substitution of $\bar x_i$ by 
$\frac{1}{M}\int_0^t{\rm d} s \ \mathcal G_+(s)G(s)$ and $\bar x(s)$ by $\frac{1}{M}\int_s^t{\rm d} u \ \mathcal G_+(u-s)G(u)$ [the first and 
the third terms of the rhs of Eq.~(\ref{eq:solxbar}) cancel].
Moreover, after the integration over $\bar x_i$ the fifth and the last terms of the rhs of Eq.~(\ref{eq:Sigma3}) cancel, too. 
Finally, the Gaussian integral over $x_i$ yields
\begin{widetext}
\begin{eqnarray}
\mathcal J[F,G] &=& 
- \frac{\Lambda'}{2\hbar M_0} \left(\int_0^t{\rm d} s\ G(s)\dot{\mathcal G}_+(s)\right)^2
+ \frac{M_0}{4\hbar\epsilon^2 M} \bar x_m \int_0^t{\rm d} s \ \mathcal G_+(s) G(s) 
\nonumber\\
&&
- \frac{\Lambda'}{\hbar M M_0}\int_0^t{\rm d} s\ G(s)\dot{\mathcal G}_+(s) \int_0^t{\rm d} s'\int_0^{s'}{\rm d} u' 
\mathcal G_+(s'-u')C_1(u')G(s') 
-\frac{\Omega' M_0}{2\hbar M^2}\left(\int_0^t{\rm d} s\ G(s) \mathcal G_+(s)\right)^2 
\nonumber\\
&&
+ \frac{i}{\hbar M} \int_0^t{\rm d} s\int_0^s{\rm d} u \ \mathcal G_+(s-u)G(s)F(u) 
+\frac{1}{\hbar M^2}\int_0^t{\rm d} s \ \mathcal G_+(s) G(s)\int_0^t{\rm d} s'\int_{s'}^t{\rm d} u' 
\mathcal G_+(u'-s')C_2(s')G(u') 
\nonumber\\
&&
+ \frac{ i\Lambda'}{\hbar\epsilon^2}x_m\left[\int_0^t{\rm d} s \ G(s)\dot{\mathcal G}_+(s) 
+\frac{1}{M}\int_0^t{\rm d} s\int_0^s {\rm d} u\ \mathcal G_+(s-u)G(s)C_1(u)\right] 
\nonumber\\
&&
- \frac{1}{2\hbar M^2 M_0}\int_0^t{\rm d} s\int_0^t{\rm d} s'\int_s^t{\rm d} u\int_{s'}^t{\rm d} u' \mathcal G_+(u-s) R''(s,s') 
\mathcal G_+(u'-s') G(u) G(u') \; .
\label{eq:Jfinal}
\end{eqnarray}
\end{widetext}
Here we introduced the kernel
\begin{eqnarray}\label{eq:def10}
R''(s,s') = \Lambda' C_1(s)C_1(s') + R'(s,s')\; ,
\end{eqnarray}
where we used Eq.~(\ref{eq:KR}). Equation~(\ref{eq:Jfinal}) is the central result of the first part of this paper. 
It allows us to derive all non-equilibrium correlation functions in a systematic way. Direct applications 
of this method will be presented in Secs.~\ref{noneq} and \ref{luttinger}.


\subsection{The correlation function}

The two-time correlation function Eq.~(\ref{eq:cf}) has two contributions,
\begin{align}
\langle \hat q(t) \hat q(t') \rangle &= \frac{1}{2}\langle [\hat q(t), \hat q(t')]_+ \rangle + \frac{1}{2}\langle [\hat q(t), \hat q(t')]_- \rangle\nonumber\\
&= \mathcal C(t,t') + i \mathcal A(t-t') 
\; .
\end{align}
The first term, the average of the anti-commutator or symmetrized contribution, is real and the second one, 
the average of the commutator or anti-symmetrized contribution, is imaginary and proportional to the linear response function, 
${\mathcal R}(t,t')$,
as shown by the Kubo formula. For generic Gaussian initial conditions, Eq.~(\ref{eq:proj}),  one finds
\begin{widetext}
\begin{eqnarray}
\label{eq:corrgen}
\mathcal C(t,t') &=&  
\frac{\Lambda'\hbar}{M_0}
\left[
\dot{\mathcal G}_+(t)\dot{\mathcal G}_+(t') 
+ \frac{\dot{\mathcal G}_+(t)}{M}\int_0^{t'}{\rm d} u \ \mathcal G_+(t'-u)C_1(u)
+ \frac{1}{M}\dot{\mathcal G}_+(t') \int_0^t{\rm d} u \ \mathcal G_+(t-u)C_1(u) \right] 
\nonumber\\
&&
+\frac{\hbar \Omega' M_0}{M^2} \mathcal G_+(t) \mathcal G_+(t') 
- \frac{\hbar}{M^2}\left[\mathcal G_+(t)\int_0^{t'} \ \mathcal G_+(t'-u)C_2(u)
+ \mathcal G_+(t')\int_0^{t} \ \mathcal G_+(t-u)C_2(u) \right] 
\nonumber\\
&&
+\frac{{\Lambda'}^2}{\epsilon^4}x_m^2\left[\dot{\mathcal G}_+(t)+\frac{1}{M}\int_0^t{\rm d} u \ \mathcal G_+(t-u)C_1(u)\right]
\left[\dot{\mathcal G}_+(t')+\frac{1}{M}\int_0^{t'}{\rm d} u \ \mathcal G_+(t'-u)C_1(u)\right] 
\nonumber\\
&&
-\frac{M_0^2}{16\epsilon^4 M^2}\bar x_m^2 \mathcal G_+(t)\mathcal G_+(t') 
+ \frac{\hbar}{M^2 M_0} 
\int_0^t{\rm d} u\int_0^{t'}{\rm d} u' \ \mathcal G_+(t-u) R''(u,u') \mathcal G_+(t'-u') 
\; .
\end{eqnarray}
\end{widetext} 
To represent a pure initial condition that is initially decoupled from the bath, as in Eq.~(\ref{fac}), we set $M = M_0$, 
$\Lambda' = \epsilon^2$ and $\Omega' = 1/4\epsilon^2$ as well as $C_1(s) = 0$, $C_2(s) = 0$ and $R''(u,u') = M K(u-u')$. 
A non-factorized initial state with -- say -- an initial position measurement of ``width'' $\epsilon$ is obtained with 
$x_m = q_m$ and $\bar x_m = 0$.
In the classical case the non equilibrium correlator of a Brownian particle is given by Eq.~(\ref{app:correlc}).
 
\section{Non equilibrium dynamics after quantum quenches.}\label{noneq}

Quenches from a high temperature initial state have been extensively studied in the literature. They correspond to the case of a 
factorizing density matrix as in Eq.~(\ref{eq:initialfac}). In this section we will study the non-equilibrium dynamics of a quantum Brownian 
particle after quenches from different initial states. As already mentioned in the Introduction, there are two experimental scenarios 
that are of interest to us. In the first one the initial position of a Brownian particle trapped by a harmonic potential
is measured at $t=0$. We will study this case in the first part of this section by focusing on the two-time correlation 
function.  The second scenario consists in a quench of the trapping potential, which in the case of a harmonic potential 
corresponds to an abrupt change in the trapping frequency. Such a quench will be studied in the second part of this section. 
In both cases, 
we will derive the asymptotic equilibration behavior of the system in the presence of Ohmic dissipation. 

\subsection{A particle in a harmonic potential with an initial position measurement} 
\label{sec:free}

The general results in Sec.~\ref{genfunc} are here specialized to the case of a particle trapped in a harmonic potential,
on which a position measurement is performed at $t=0$. As we work with the same particle initially and subsequently, 
$M_0=M$, while $\omega_0 = \omega>0$. At $t=0$ a measurement of the particle position is performed with outcome $q_m = 0$ 
and uncertainty $\Delta$. The initial density matrix is given by Eq.~(\ref{eq:init3}). Note that the particle is 
permanently coupled to the bath, 
hence the initial density matrix does not factorize. 
Thus, our starting point is the general expression Eq.~(\ref{eq:corrgen}) with $M = M_0$, $\omega=\omega_0$ and $x_m = \bar x_m =0$. 
The Laplace transform of the correlator  reads
\begin{widetext}
\begin{eqnarray}
\label{corr}
\tilde{\mathcal C}(\lambda,\kappa) = \frac{\hbar}{M}\tilde{\mathcal G}_+(\lambda)\tilde{\mathcal G}_+(\kappa)
\left\{
  \Lambda'\lambda\kappa + \frac{\Lambda'}{M} \lambda \tilde C_1(\kappa) + \frac{\Lambda'}{M} \kappa \tilde C_1(\lambda) 
+ \frac{\Omega' M_0^2}{M^2} 
- \frac{1}{M}\tilde C_2(\lambda) - \frac{1}{M}\tilde C_2(\kappa) + \frac{1}{M^2}\tilde{R''}(\lambda,\kappa) \right\}\; ,  
\end{eqnarray}
\end{widetext}
an expression that can be simplified by using the method explained in App.~\ref{sec:FDT}.

Introducing the function $\tilde h_k(\lambda) = \tilde g_k(\lambda)/M + \lambda$,  Eq.~(\ref{corr}) can be written as
\begin{align}
\tilde {\mathcal C}(\lambda,\kappa) &= \frac{\hbar}{M}\tilde{\mathcal G}_+(\lambda)\tilde{\mathcal G}_+(\kappa)\frac{(\Lambda'-\Lambda)}{(\beta\hbar\Lambda)^2}
\sum_{k,k'} u_k u_{k'} \tilde h_k(\lambda) \tilde h_{k'}(\kappa) \nonumber\\
&+ \frac{\hbar}{4M\epsilon^2}\tilde{\mathcal G}_+(\lambda)\tilde{\mathcal G}_+(\kappa) + \frac{\tilde{\mathcal C}^{\rm 1eq}(\lambda) + \tilde{\mathcal C}^{\rm 1eq}(\kappa)}{\lambda + \kappa}\; ,
\end{align}
with the equilibrium correlation function $\tilde {\mathcal C}^{\rm 1eq}(\lambda)$ in the Laplace domain 
defined in Eq.~(\ref{eq:corrL}). In App.~\ref{sec:FDT} it is shown that $\tilde h_k(\lambda)$ can be written in terms 
of $\tilde{\mathcal G}_+^{-1}(\lambda)$ and $\tilde{\mathcal G}_+^{-1}(|\nu_k|)$ [see Eq.~(\ref{eq:hk})]. For $\omega = \omega_0$ we use 
the fact that $u_k =  \tilde{\mathcal G}_+(|\nu_k|)$ and we find from the definition of the equilibrium correlator 
\begin{eqnarray}
\tilde{\mathcal C}^{\rm 1eq}(\lambda) = \frac{1}{\beta M}\sum_k \frac{\lambda}{\nu_k^2-\lambda^2}
\left[\tilde{\mathcal G}_+(\lambda) - \tilde{\mathcal G}_+(|\nu_k|)\right] \;,
\end{eqnarray}
which is derived in App.~\ref{sec:FDT} [see Eq.~(\ref{eq:corrL})], that the desired non equilibrium correlation function of a quantum 
Brownian particle with initial position measurement reads
\begin{eqnarray}
\label{eq:correl1}
\tilde{\mathcal C}(\lambda,\kappa) 
&=& 
\displaystyle{\frac{M}{\hbar}\frac{\Lambda'-\Lambda}{\Lambda^2}
\tilde{\mathcal C}^{\rm 1eq}(\lambda)\tilde{\mathcal C}^{\rm 1eq}(\kappa) + 
\frac{\hbar}{4M\epsilon^2}\tilde{\mathcal G}_+(\lambda)\tilde{\mathcal G}_+(\kappa)} 
\nonumber \\
&&
+ \displaystyle{\frac{\tilde{\mathcal C}^{\rm 1eq}(\lambda) + \tilde{\mathcal C}^{\rm 1eq}(\kappa)}{\lambda + \kappa}} \; .
\end{eqnarray}
The classical correlator for an initial position measurement with outcome $q^0 = 0$ and uncertainty $\Delta$ can be obtained from Eq.~(\ref{app:corrclassical3}) by replacing $\langle {q^0}^2 \rangle$ by $\Delta^2$. The initial momentum is not measured and it is therefore distributed according to the Boltzmann--law with $\langle v_0^2\rangle = (\beta M)^{-1}$. In the limit of a sharp position measurement ($\Delta\to 0$) the classical correlator of an equilibrium particle reads 
$\tilde {\mathcal C}^{\rm eq}(\lambda,\kappa) - \beta M\omega^2 \ \tilde{\mathcal C}^{\rm 1eq}(\lambda)\tilde{\mathcal C}^{\rm 1eq}(\kappa)$. 
As to the quantum correlator we note that $\Lambda' = 0$ for $\Delta \to 0$ and the sum of the first and the third term in the 
rhs of Eq.~(\ref{eq:correl1}) yields 
$\tilde {\mathcal C}^{\rm eq}(\lambda,\kappa) - M/(\hbar\Lambda) \ \tilde{\mathcal C}^{\rm 1eq}(\lambda)\tilde{\mathcal C}^{\rm 1eq}(\kappa)$ which already has the form of its classical counterpart. 
It is easy to show that in the high temperature limit $\beta\hbar \ll |\omega^2 - \gamma^2/4|$ the two expressions coincide exactly. The role of the second term in the rhs of Eq.~(\ref{eq:correl1}) remains to be discussed: it describes the initial momentum due to Heisenberg's uncertainty relation. Consequently, it diverges when the initial position measurement becomes sharp unless one considers that $\hbar/(M\epsilon^2) = \hbar^2/(M^2 \Delta^2) \to 0$ even though $\Delta \to 0$. 
More precisely, when $\beta\hbar |\omega^2 - \gamma^2/4|^{1/2} \ll 1$ we have $\mathcal C^{\rm 1eq} \sim 1/(\beta M \omega^2)$ and $\mathcal G_+ \sim 1/\omega$. In order for the second term of the rhs of Eq.~(\ref{eq:correl1}) to be small compared to the other terms the condition $\Delta \gg \lambda_T$ must hold, with $\lambda_T = \sqrt{\beta\hbar^2/(2\pi M)}$ the thermal de Broglie--wavelength of the particle. Only then can one speak of a classical particle: the condition $\beta\hbar|\omega^2 - \gamma^2/4|^{1/2}\ll 1$ that properly defines the ``high temperature regime'' is not sufficient. One also has to take the ``macroscopic measurement limit'' defined through $\Delta \gg \lambda_T$.

In the real time domain the quantum correlation function is
\begin{eqnarray}
\label{eq:corrinitialmeasure}
\mathcal C(t,t') &=& 
\frac{M}{\hbar}\frac{\Lambda'-\Lambda}{\Lambda^2}\mathcal C^{\rm 1eq}(t')
\mathcal C^{\rm 1eq}(t')
 \nonumber \\
&&+ \frac{\hbar}{4M\epsilon^2}\mathcal G_+(t)\mathcal G_+(t') 
+ \mathcal C^{\rm 1eq}(|t-t'|) \; ,
\end{eqnarray}
where $\mathcal C^{\rm 1eq}(t)$ is the real time equilibrium correlation function. This result applies to any kind of spectral density 
of the bath. We will use the correlation function Eq.~(\ref{eq:corrinitialmeasure}) in Sec.~\ref{luttinger} to study the non equilibrium 
dynamics of an impurity in a Luttinger liquid bath for which a specific spectral density of the bath applies.

We are interested in the equilibration behavior of Eq.~(\ref{eq:corrinitialmeasure}) in the presence of Ohmic dissipation for which the spectral density behaves as $S(\omega) \sim \gamma \omega$ for small $\omega$, with $\gamma$ playing the role of a friction coefficient [for more details see App.~\ref{app:longtime}]. It is of special interest to study the strong quantum regime 
$\beta\hbar\gg |\omega^2 - \gamma^2/4|^{-1/2}$. 
By using
 the long--time limits Eq.~(\ref{eq:longC}) and Eq.~(\ref{eq:longG}) presented in App.~\ref{app:longtime} valid for $t \gg \gamma^{-1}$ we find 
 that the propagator $\mathcal G_+(t)$ exponentially approaches zero whereas the equilibrium correlation function 
 $\mathcal C^{\rm 1eq}(t)$ relaxes with a power law  $\sim t^{-2}$ for $t\to\infty$. In the long--time limit the correlator 
 Eq.~(\ref{eq:corrinitialmeasure}) thus relaxes as fast as the squared equilibrium correlation function. Therefore, for 
 $\beta\hbar \gg |\omega^2 - \gamma^2/4|^{-1/2}$~:
\begin{align}
\mathcal C(t,t') &= \mathcal C^{\rm 1eq}(|t-t'|) + \mathcal O\left[(t t')^{-2}\right]  
\end{align}
when $ t,t' \gg \gamma^{-1}$, 
and the equilibrium function $\mathcal C^{\rm 1eq}(|t-t'|) = \mathcal C^{\rm eq}(t,t')$ is asymptotically approached during the algebraic 
relaxation of the non--equilibrium terms. Consequently, care has to be taken in experiments when an equilibrium system is desired 
at very low temperatures after an initial position measurement. The relaxation of the system is slow independently  of 
the dissipation strength $\gamma$. At high temperatures $\beta\hbar \ll |\omega^2 - \gamma^2/4|^{-1/2}$ the relaxation is of order $\mathcal O(e^{-\gamma t})$ [see the 
discussion in App.~\ref{app:longtime}] and therefore exponential as in the classical limit.

\subsection{Quantum quench in the confining potential.}\label{sec:nonfree}

In this section we desire to gain insight into the equilibration process of a quantum Brownian particle after an abrupt change in 
the trapping frequency. At $t<0$ the particle is confined in a harmonic potential with frequency $\omega_0 > 0$. At $t=0$ the 
experimentalist abruptly changes the strength of the harmonic potential resulting in a higher or lower trapping 
frequency.
We do not consider an initial position measurement since we assume that the particle is already localized by the initial harmonic 
trap. Hence we set $\Lambda = \Lambda'$, $\Omega = \Omega'$, $x_m = \bar x_m = 0$ and $\epsilon \to \infty$. By starting 
from Eq.~(\ref{corr}) and by using Eq.~(\ref{Rtilde}) it is straightforward to show with the methods employed in App.~\ref{sec:FDT} that the 
correlator in the Laplace domain reads
\begin{align}\label{eq:corrquench}
\tilde{\mathcal C}(\lambda,\kappa) = \frac{\hbar}{M} \frac{\tilde{\mathcal G}_+(\lambda) \tilde{\mathcal G}_+(\kappa)}{\tilde{\mathcal G}_+^0(\lambda) \tilde{\mathcal G}_+^0(\kappa)} \frac{1}{\lambda+\kappa}\left[ \tilde{\mathcal C}^{\rm 1eq}_0(\lambda) + \tilde{\mathcal C}^{\rm 1eq}_0(\kappa) \right]\; ,
\end{align} 
where $\tilde{\mathcal G}_+^0(\lambda) = 1/[\lambda^2 + \lambda \tilde\gamma(\lambda) + \omega_0^2]$ is the propagator with initial frequency $\omega_0$, and 
\begin{align}
\tilde{\mathcal  C}^{\rm 1eq}_0(\lambda) = \frac{1}{\beta M} \sum_k \frac{\lambda}{\nu_k^2 - \lambda^2} \left[\tilde{\mathcal G}_+^0(\lambda) - \tilde{\mathcal G}_+^0(|\nu_k|)\right]
\end{align}
is the equilibrium correlation function of a particle in a harmonic potential with frequency $\omega_0$ [see Eq.~(\ref{eq:corrL})]. 
The structure of Eq.~(\ref{eq:corrquench}) is very different from the classical counterpart Eq.~(\ref{app:corrclassical2}). 
Still, by using the high temperature approximation 
$\tilde{\mathcal C}_0^{\rm 1eq}(\lambda) \simeq -\left[\tilde{\mathcal G}_+(\lambda) - 1/\omega_0^2\right]/\lambda$ 
[see the end of the discussion in App.~\ref{app:longtime}] one recovers the classical expression Eq.~(\ref{app:correlc}).

The equilibration time for Ohmic dissipation in the strong quantum regime $\beta\hbar \gg |\omega^2 - \gamma^2/4|^{-1/2}$ 
can be found in the following way. Note first that when $\tilde{\mathcal C}_0^{\rm 1eq}(\lambda)$ is multiplied by $\tilde{\mathcal G}_+(\lambda)/\tilde{\mathcal G}_+^0(\lambda)$ one obtains a new function that we call $\tilde{\mathcal C}'$:
\begin{eqnarray}
\label{eq:Cprime}
\tilde{\mathcal C}'(\lambda) 
&=& \frac{1}{\beta M} \sum_k \frac{\lambda}{\nu_k^2 - \lambda^2} \nonumber\\
&& 
\times \left[\tilde{\mathcal G}_+(\lambda) 
- \frac{\tilde{\mathcal G}_+(\lambda)\tilde{\mathcal G}^0_+(|\nu_k|)}
{\tilde{\mathcal G}_+^0(\lambda)\tilde{\mathcal G}_+(|\nu_k|)}
\tilde{\mathcal G}_+(|\nu_k|)\right] \; .
\end{eqnarray}
Second, we observe that the Laplace transform of 
$(\partial_t + \partial_{t'})f(t,t')$ is equal to $(\lambda+\kappa)\tilde f(\lambda,\kappa) - \tilde f(t=0;\kappa) - \tilde f(\lambda;t'=0)$ 
where $\tilde f(\lambda,\kappa)$ is the Laplace transform of the generic function $f(t,t')$ with respect to both $t$ and $t'$ 
while $\tilde f(\lambda;t'=0)$ [$\tilde f(t=0;\kappa)$] is the Laplace transform of $f(t,t'=0)$ [$f(t=0,t')$] 
with respect to $t$ ($t'$).
Now, by choosing $f(t,t') = \mathcal C(t,t')$ we find for the correlation function Eq.~(\ref{eq:corrquench}) in the time domain
\begin{eqnarray}
\label{eq:eq10}
(\partial_t + \partial_{t'})\mathcal C(t,t') 
&=& \mathcal C'(t)\left[1+(\omega^2-\omega^2_0)\mathcal G_+(t')\right] 
\nonumber\\
&&
+ \mathcal C'(t')\left[1+(\omega^2-\omega^2_0)\mathcal G_+(t)\right] 
\nonumber\\
&&
- \mathcal C(t,0) - \mathcal C(0,t')\; .
\end{eqnarray}
From Eq.~(\ref{eq:corrquench}) we easily find the expression of the Laplace transform of ${\mathcal C}(t,t'=0)$ by multiplying 
$\tilde {\mathcal C}(\lambda,\kappa)$ by $\kappa$ and by taking the limit $k\to\infty$:
\begin{eqnarray}
\tilde{\mathcal C}(\lambda;t' = 0) &=& \tilde{\mathcal C}_0^{\rm 1eq}(\lambda)
\left[1+(\omega^2-\omega^2_0)\tilde{\mathcal G}_+(\lambda)\right] 
\nonumber\\
&=&
 \tilde{\mathcal C}'(\lambda) \; .
\end{eqnarray}
Accordingly, Eq.~(\ref{eq:eq10}) simplifies to 
\begin{eqnarray}\label{eq:eq11}
(\partial_t + \partial_{t'}) \mathcal C(t,t') &=& \mathcal C'(t)(\omega^2-\omega^2_0)\mathcal G_+(t') \nonumber\\
&&+ \mathcal C'(t')(\omega^2-\omega^2_0)\mathcal G_+(t) \; .
\end{eqnarray}
Finally, it is clear that $(\partial_t + \partial_{t'}) \mathcal C(t,t') = 0$ when $\mathcal C(t,t') = 
\mathcal C^{\rm eq}(t,t') = \mathcal C^{\rm 1eq}(|t-t'|)$ so that the terms of the rhs of Eq.~(\ref{eq:eq11}) can be understood 
as the derivative of the non equilibrium part of $\mathcal C(t,t')$. In combination with the results of the discussion in 
App.~\ref{app:longtime} the equilibration behavior of the correlator at low temperatures after a quench in the trapping potential 
can be summarized as follows [by noting that the asymptotic long--time behavior of $\mathcal C'(t)$ and $\mathcal C^{\rm 1eq}(t)$ are 
identical]:
\begin{align}
\mathcal C(t,t') = \mathcal C^{\rm 1eq}(|t-t'|) &+ \mathcal O\left[e^{-\gamma t/2} / t' + e^{-\gamma t'/2} / t\right]
\end{align} 
for $t,t' \gg \gamma^{-1}$.
Consequently, the non equilibrium contributions are exponentially suppressed which leads to a faster equilibration than the one 
found after an initial position measurement.

At high temperatures the relaxation of $\mathcal C'(t)$ and $\mathcal C^{\rm 1eq}(t)$ are both exponential of 
$\mathcal O(e^{-\gamma t/2})$ so that in the high temperature regime $\mathcal C(t,t') = \mathcal C^{\rm 1eq}(|t-t'|) + 
\mathcal O\left[e^{-\gamma (t+t')/2}\right]$.

\section{Evolution of an impurity in a 1D Bose gas}\label{luttinger}

Many 1D systems are described by the Tomonaga-Luttinger theory \cite{Tomo50, Lutt63, MattisLieb65}. 
The evolution of impurities in such a Luttinger liquid (LL) has attracted much attention in recent years \cite{Giamarchi07, Johnson12} 
since this problem can be explicitly realized with cold atom systems. In particular, modern techniques allow one 
to tune the interspecies interaction strength~\cite{Thal08,Peano05,Olsh98} so that it has become 
possible to study the diffusion of a minority species within an ensemble of 
majority atoms, as a function of the interaction and the trapping potential \cite{Giamarchi11}. 

In this part of our work we apply the non equilibrium formalism developed in the first part of our article to such an impurity--LL system. In 
particular, we seek to mimic the experimental process described in~\cite{Giamarchi11} with our theoretical description. In this 
experiment the impurity atom is trapped in a 1D harmonic potential together with an ensemble of a different kind 
of atoms that form the LL. 
The impurity is initially localized  at the center of the confining potential by a laser blade. When the whole impurity--LL system 
reaches equilibrium the impurity is released. The equal-times position correlation function of the impurity, $\mathcal C(t,t)$, 
then shows damped oscillations which strongly suggest that the impurity is \emph{de facto} a quantum Brownian particle moving in a 
quantum liquid bath. 

In the following we will present a precise description of the impurity motion in the LL from the quantum Brownian motion point of view. 
The LL itself will play the role of an exotic quantum bath that we here characterize.

\subsection{The impurity model}

The impurity and the atoms constituting the bath are all  confined in a harmonic potential. 
We therefore take the Hamiltonian of the  impurity, $\hat{\mathcal H}_S$, to be of the standard form Eq.~(\ref{eq:Hs}) without 
external force ($H=0$) and with the harmonic potential 
$V(\hat q) = \frac{M\omega^2}{2} \hat q^2$. We assume that the interaction Hamiltonian between the position operator 
$\hat q$ of the impurity and the density of the boson liquid is of the form
\begin{eqnarray}\label{HLcoupling}
\displaystyle{
\hat{\mathcal H}_{SB} = \int{\rm d} x {\rm d} y\ U(x-y)\hat\rho(y)\delta(x-\hat q)} 
 \; ,
\end{eqnarray}
with the density operator $\hat\rho(x)$ of the LL approximately described by
\begin{eqnarray}
\hat\rho(x) \simeq \rho_0(x) - \frac{1}{\pi}\nabla\hat\phi(x) 
 \;,
\end{eqnarray}
where $\rho_0(x)$ is the unperturbed density of the fluid in the 1D trap and 
$\hat\phi(x)$ is the density variation. The Hamiltonian of the free LL reads
\begin{eqnarray}\label{HLfree}
\hat{\mathcal H}_B = \frac{\hbar}{2\pi}\int{\rm d} x\left[\frac{u K_L}{\hbar^2}(\pi\hat\Pi(x))^2+\frac{u}{K_L}(\nabla\hat\phi(x))^2\right] \; ,
\end{eqnarray}
where $\hat\Pi(x)$ and $\hat\phi(x)$ are conjugate operator fields. Equation~(\ref{HLfree}) describes the low-energy properties 
of a Lieb-Liniger gas~\cite{Lieb-Liniger} 
with a contact interaction potential $\hbar v_L \delta(x)$. The parameters $u$ (with the dimension of a velocity) 
and $K_L$ (dimensionless) have to 
be determined numerically for general $v_L$. 
In order to reduce the complexity of the problem we will assume the background density $\rho_0$ to be constant in the following. Accordingly, we 
define
\begin{eqnarray}
\rho_0 \equiv \frac{1}{L}\int{\rm d} x \ \rho_0(x) \; ,
\end{eqnarray}
where $L$ is a length scale of the order of the length of the trap. Note that in this modeling we have not added the 
quadratic confining potential to the LL.

Since the Bose gas is confined in a space of length $L$, the wave vectors are quantized with values $k_n = \pi n/L$ with $n$ an integer. 
The Fourier representation of Eq.~(\ref{HLcoupling}) is
\begin{eqnarray}
\hat{\mathcal H}_{SB} = \frac{1}{\sqrt{L}} \sum_n \tilde U_{k_n} \ e^{ik_n\hat q}\left[ - \frac{ik_n}{\pi}\tilde\phi(k_n)\right] \; ,
\end{eqnarray}
where we used
$\delta(x-x') = (1/L)\sum_n e^{ik_n(x-x')}$ and we neglected a constant contribution.
We assume that the potential $U$ has the form
\begin{eqnarray}
\tilde U_k = \hbar v e^{-|k|/k_c} \; ,
\end{eqnarray}
with some finite cutoff $k_c$ that depends on the microscopic properties of the interaction. The parameter $v$ has the dimension 
of a velocity and it determines the strength of the impurity--bath interaction potential. 

After redefining the fields according to
\begin{eqnarray}
\hat\phi(x) &\mapsto& \sqrt{(\pi K_L/\hbar)} \ \hat\phi(x) 
\end{eqnarray}
and 
\begin{eqnarray}
\hat\Pi(x) &\mapsto& \sqrt{(\hbar/\pi K_L)} \  \hat\Pi(x) \; ,
\end{eqnarray}
one introduces the bosonic ladder operators
\begin{align}
\hat b_k &= \sqrt{\frac{|k|}{2\hbar}}\left(\tilde\phi(k)+\frac{i}{|k|}\tilde\Pi(k)\right)
\end{align}
and 
\begin{align} 
{\hat b}_{-k}^\dagger &= \sqrt{\frac{|k|}{2\hbar}}\left(\tilde\phi(k)-\frac{i}{|k|}\tilde\Pi(k)\right)
\; , 
\end{align}
which describe bosonic density wave excitations with sound velocity $u$ \cite{Giamarchi03}. 
The full Hamiltonian now takes the form of  the Fr\"ohlich polaron Hamiltonian, 
which in the second quantization language reads
\begin{widetext}
\begin{align}\label{eq:fullLLham}
\hat{\mathcal H} = \sum_{k\in\{k_n\}} \hbar u|k| \hat b_k^\dagger\hat b_k
 + \frac{\hat p^2}{2 M} + \frac{M \omega^2}{2} \hat q^2
 - \frac{1}{\sqrt{L}}\sum_k\left(\frac{K_L}{2\pi|k|}\right)^{1/2}\tilde U_k \left[(ik e^{ik\hat q})\ \hat b^\dagger_k + (ik e^{ik\hat q})^*\ \hat b_k\right]
\; .
\end{align}
\end{widetext}
For each $k$ mode the coupling between the operator 
$e^{ik\hat q}$ and the bath operators $\hat b^\dagger_k$ and $\hat b_k$ is bilinear, so we can use the general results derived in 
Sec.~\ref{roleinit} by considering that $e^{ik\hat q}$ represents a different operator for each $k$ that is coupled to one harmonic oscillator. The resulting influence functional is hence a sum of one-particle--one-oscillator influence functionals. By 
using Eq.~(\ref{eq:phi1}) we thus find 
\begin{widetext}
\begin{eqnarray} 
\label{phi1L}
\Phi[q^+,q^-,q^0] = 
-
\sum_k && \;\;
\left\{
\int_0^{\beta\hbar}{\rm d}\tau\int_0^\tau{\rm d}\sigma \ K_k(-i\tau+i\sigma)e^{ik q^0(\tau)}e^{-ik q(\sigma)} 
\right.
\nonumber\\
&&
+ i\int_0^{\beta\hbar}{\rm d}\tau\int_0^t{\rm d} s \ K_{-k}^*(s-i\tau)e^{ik q^0(\tau)}
\left[e^{-ik q^+(s)}-e^{-ik q^-(s)}\right] 
\nonumber\\
&&
\left.
- \int_0^t{\rm d} s\int_0^s{\rm d} u \left[e^{ik q^+(s)}-e^{ik q^-(s)}\right]
  \left[K_{-k}(s-u)e^{-ik q^+(u)}-K_{-k}^*(s-u)e^{-ik q^-(u)}\right] 
   \right\}
  \; . 
\end{eqnarray}
\end{widetext}
Note that, since in the initial Hamiltonian there is no counter-term the counter-terms in Eq.~(\ref{eq:phi1}) 
are absent in Eq.~(\ref{phi1L}). The $k$-dependent kernel reads
\begin{eqnarray}
K_k(\theta) = \frac{1}{L} \frac{K_L|k|}{2\pi \hbar} \ |\tilde U_k|^2 \ \frac{\cosh[u|k|\beta\hbar/2-i\theta u|k|]}{\sinh[u|k|\beta\hbar/2]}\; .
\end{eqnarray}

A well known feature of the Fr\"ohlich polaron Hamiltonian (\ref{eq:fullLLham}) is its polaronic mass shift which leads 
to an effective impurity mass $M^*>M$ greater than the ``bare mass''~\cite{We08}. 
Another process described by this Hamiltonian is that during a collision between the impurity and an LL atom
the former loses momentum $\hbar k$ by creating a density wave excitation $b^\dagger_k$ in the LL.
However, the LL is itself confined in a harmonic potential and one may expect to have some momentum transfer 
absorbed (or provided) from the LL to the optical trap. We will here simply assume that 
the spring constant of the optical potential is renormalized by such a process in such a way that it balances the 
polaronic mass shift and we will henceforth work with the bare Hamiltonian (\ref{eq:fullLLham}). 

If the oscillations are small (if the impurity potential is sufficiently steep) we can expand the $e^{i k \hat q}$ in Eq.~(\ref{phi1L}) 
to second order in $k$. 
Note that the linear order vanishes in Eq.~(\ref{phi1L}) due to the symmetry $k\mapsto -k$. 
Ignoring the zero-th order in $k$ we can make the replacements
\begin{align}
\label{mapsto}
&e^{ikq^0(\tau)}e^{-ikq^0(\sigma)} \mapsto -\frac{k^2}{2}\left[q^0(\tau) - q^0(\sigma)\right]^2 \; , \nonumber\\
&e^{ikq^0(\tau)} \! \left[e^{-ikq^+(s)} - e^{-ikq^-(s)}\right] \mapsto \nonumber\\
&k^2 q^0(\tau) \ \left[q^+(s) - q^-(s)\right] - \frac{k^2}{2} \left[q^+(s)^2 - q^-(s)^2\right] 
\; ,
\nonumber \\
&\left[e^{ik q^+(s)} - e^{ik q^-(s)}\right]\times\\
&\left[K_{-k}(s-u)e^{-ikq^+(u)} - K_{-k}^*(s-u)e^{-ikq^-(u)}\right] \mapsto 
\nonumber\\
&k^2\left[q^+(s)-q^-(s)\right]\left[K_k(s-u)q^+(u) - K_k^*(s-u)q^-(u)\right] 
\nonumber\\
&-\frac{k^2}{2}\left[q^+(s)^2-q^-(s)^2\right]\left(K_k(s-u)-K_k^*(s-u)\right) \nonumber \;.
\end{align}

\subsection{Steep potential: The Luttinger bath and the initial condition.}

If the external potential is steep we expect the Gaussian approximation of the Fr\"ohlich Hamiltonian (\ref{mapsto}) to be valid. Up to this order in $k$ each $k$--mode plays the role of one bath harmonic oscillator. Accordingly, for $L \to \infty$ we define 
the spectral density as
\begin{align}\label{spectralL}
S(\omega) &= \frac{\pi}{L}\sum_{k\in\{k_n\}} \frac{K_L |k|^3}{2\pi\hbar} \ |\tilde U_k|^2 \ \delta(\omega - u|k|) \nonumber\\
&\simeq \frac{K_L\omega^3\hbar v^2}{\pi u^4}e^{-\omega/\omega_c} 
= \frac{\pi\mu}{4}\left(\frac{\omega}{\omega_c}\right)^3 e^{-\omega/\omega_c}
\; , 
\end{align}
with $\omega_c = u|k_c|/2$ and 
\begin{equation}
 \mu = \frac{4\hbar K_L v^2\omega_c^3}{\pi^2u^4} \;  .
\end{equation} 
Hence, the bath induces \emph{super--Ohmic} dissipation with a power law behavior $S(\omega) \sim \omega^3$ 
for small $\omega$. This is the main difference from the analysis presented in Sec.~\ref{genfunc}.

In terms of the fundamental kernel [c.f. Eq.~(\ref{app:K})]
\begin{eqnarray}
K(\theta) = \int_0^\infty\frac{{\rm d}\omega}{\pi}S(\omega) \frac{\cosh[\omega(\beta\hbar/2-i\theta)]}{\sinh[\omega\beta\hbar/2]} 
\end{eqnarray} 
and with
$\int_0^{\beta\hbar}{\rm d}\tau K^*(s-i\tau) = M\gamma(s)$
and
$\int_0^s{\rm d} u \left[K(s-u)-K^*(s-u)\right] = -iM\gamma(0) + i M \gamma(s)$, 
where $\gamma(s)$ is defined in Eq.~(\ref{eq:gamma}), we obtain
\begin{widetext}
\begin{eqnarray}
\label{phi3L}
&& \Phi[q^+,q^-,q^0] 
=
\frac{1}{4}\int_0^{\beta\hbar}{\rm d}\tau{\rm d}\sigma \ K(-i\tau+i\sigma)\left[q^0(\tau) - q^0(\sigma)\right]^2 
-i\int_0^{\beta\hbar}{\rm d} \tau \int_0^t{\rm d} s \ K^*(s-i\tau) q^0(\tau) \left[q^+(s) - q^-(s)\right]
\nonumber\\
&& 
\qquad
+\int_0^t{\rm d} s\int_0^s{\rm d} u \left[q^+(s)-q^-(s)\right]\left[K(s-u)q^+(u)
- K^*(s-u)q^-(u)\right] 
+ \frac{iM\gamma(0)}{2}\int_0^t{\rm d} s\left[q^+(s)^2-q^-(s)^2\right]\; .
\end{eqnarray}
\end{widetext}
The first line in Eq.~(\ref{phi3L}) shows that only the non-local part of $K(-i\tau+i\sigma)$ contributes. Hence we can replace the first line 
by $\frac{1}{2}\int_0^{\beta\hbar}{\rm d}\tau{\rm d}\sigma \ k(-i\tau+i\sigma)q^0(\tau) q^0(\sigma)$. Furthermore, we see that the last line 
in Eq.~(\ref{phi3L}) exactly represents the counter-term proportional to $\mu$. By rewriting Eq.~(\ref{phi3L}) in terms of the kernels $K_R$ 
and 
$\gamma$ [see Eqs.~(\ref{app:KR}) and~(\ref{eq:gamma})] and using the transformed variables 
$x = (q^++q^-)/2$ and $\bar x = q^+-q^-$ the action becomes
\begin{widetext}
\begin{eqnarray}\label{phi4L}
&&\Sigma[x,\bar x,q^0,x_i,x_f,\bar x_i] 
= i\int_0^{\beta\hbar}{\rm d}\tau\left[ \frac{M}{2}{{\dot q}^0(\tau)}^2+\frac{M\omega^2}{2}{q^0(\tau)}^2 
+\frac{1}{2} \int_0^{\beta\hbar} \!\! {\rm d}\sigma \ k(\tau-\sigma)q^0(\tau)q^0(\sigma)\right] 
\nonumber\\
&&
\qquad
+ \int_0^{\beta\hbar} \! {\rm d}\tau\int_0^t \!{\rm d} s \ K^*(s-i\tau)q^0(\tau)\bar x(s) 
+ \int_0^t \! {\rm d} s\left[ M\dot{\bar x}(s) \dot x(s) - 
M\omega^2 \bar x(s) x(s) - M\bar x(s) \frac{{\rm d}}{{\rm d} s}\int_0^s \! {\rm d} u \ \gamma(s-u) x(u)\right] 
\nonumber\\
&&
\qquad
+ \frac{i}{2}\int_0^t {\rm d} s \int_0^t {\rm d} u\ K_R(s-u)\bar x(s) x(u) + \frac{i\hbar}{2\Delta^2}\left[x_i^2+\frac{\bar x_i^2}{4}\right]
\; .
\end{eqnarray}
\end{widetext}

We remind the reader of the main approximations used so far. First, we used the Gaussian approximation (\ref{mapsto}) of the 
Fr\"ohlich Hamiltonian. Second, in order for the action (\ref{phi4L}) to make sense we assumed $\omega$ to be large enough, 
so that non Gaussian effects are not too important. Note, that the mass and the potential renormalization can modify the oscillation amplitude and the final width of the impurity position. 
Finally, we interpreted the laser blade that initially localizes the impurity at the center of the quantum liquid as an initial 
position measurement 
with outcome $q_m = 0$ and uncertainty $\Delta$. The effect of the initial position measurement is incorporated into the action via 
the last term of the rhs of Eq.~(\ref{phi4L}). If the localization is performed itself by a very steep trapping potential with frequency $\Omega_0$ the particle is in its harmonic oscillator (with respect to $\Omega_0$) ground state at initial time. We then have
\begin{align}
&\epsilon^{-2} = 2\Omega_0 \; .
\end{align}
This approximation is disputable since one could also consider the initial localization of the impurity as stemming from an initial trapping 
potential. The subsequent release of the impurity would then rather be described by a quench in the harmonic potential [see 
Sec.~\ref{sec:nonfree}]. However, in real experiments the ``high temperature'' regime $\hbar\beta\omega \ll 1$ prevails. From the 
discussion in Secs.~\ref{sec:free} and~\ref{sec:nonfree} we know that in this case the difference between the particle motion after an initial 
position measurement and the one after an initial localization due to an initial trapping potential (followed by a quench in the potential) is 
blurred. Since it is technically easier to deal with a position measurement we prefer this method to a quench in the trapping potential. Note 
that the ``high temperature'' regime is not equivalent to the classical regime, as we pointed out in Sec.~\ref{sec:free} 
as it does not fulfill the ``macroscopic measurement'' condition $\Delta \gg \lambda_T$ with $\lambda_T$ the thermal de Broglie--
wavelength of the impurity. For more details go back to  the discussion in Sec.~\ref{sec:free}.

\subsection{Signature of a Luttinger liquid bath}

A typical experimental scenario consists in holding the impurity in the center of the trap at $t<0$ and switching off the localizing 
potential at $t=0$ to let the impurity move in the residual harmonic potential. 
Let us for a moment forget about the potential renormalization and the mass shift.
By using the results found in Sec.~\ref{sec:free} we find in the limit $\epsilon \to 0$ (which corresponds to an almost perfect localization of 
the impurity at $t=0$)
\begin{eqnarray}\label{eq:corrLaplace}
\tilde{\mathcal C}(\lambda,\kappa) &\simeq& 
\frac{\hbar}{4M \epsilon^2}\tilde{\mathcal G}_+(\lambda)\tilde{\mathcal G}_+(\kappa) - 
\frac{M}{\hbar\Lambda} \tilde{\mathcal C}^{\rm 1eq}(\lambda) \tilde{\mathcal C}^{\rm 1eq}(\kappa) \nonumber\\
&& 
 + \frac{1}{\lambda+\kappa}\left[\tilde {\mathcal C}^{\rm 1eq}(\lambda) + \tilde{\mathcal C}^{\rm 1eq}(\kappa)\right] 
\; .
\end{eqnarray}
In the time domain the correlation function reads
\begin{eqnarray}
\mathcal C(t,t') &\simeq& 
\frac{\hbar}{4M \epsilon^2} \mathcal G_+(t) \mathcal G_+(t') -\frac{M}{\hbar\Lambda} \mathcal C^{\rm 1eq}(t) \mathcal C^{\rm 1eq}(t') \nonumber \\
&& + \mathcal C^{\rm 1eq}(|t-t|) 
 \; .
\end{eqnarray}
For the moment, experimental measurements focus on  
the equal-time correlation (which corresponds to the time-dependent variance of the position)
\begin{eqnarray}\label{sigma2L}
\mathcal C(t,t) \simeq \frac{\hbar}{4 M \epsilon^2} \mathcal G_+^2(t) - \frac{M}{\hbar\Lambda}\mathcal C^{\rm 1eq}(t)^2 + 
\mathcal C^{\rm 1eq}(0) 
 \; .
\end{eqnarray}
The formula Eq.~(\ref{sigma2L}) is valid for all kinds of baths and for very small polaronic effects and potential renormalization. 

In order to gain further insight into the dynamics of the impurity we need to understand the contribution of 
each of the three terms in Eq.~(\ref{sigma2L}). 

We start from the analysis of the propagator $\mathcal G_+(t)$. To determine its time-dependence we 
need the specific form of the spectral density of the LL bath (\ref{spectralL}). 
In terms of the function
\begin{eqnarray}
\label{eq:gut}
g(z) &=& \frac{1}{2}\int_0^\infty{\rm d} \zeta \ \zeta^2 \ e^{-\zeta} \ \frac{z^2}{\zeta^2+z^2}
\nonumber\\
&=& \frac{z^2}{2} - \frac{z^4}{2}\int_0^\infty{\rm d}\zeta \ \frac{e^{-\zeta}}{\zeta^2+z^2} 
\end{eqnarray}
the Laplace transform of the damping kernel $\tilde\gamma(\lambda)$ [see Eqs.~(\ref{eq:gamma})] can be 
recast as
\begin{eqnarray}
\lambda\tilde\gamma(\lambda) = \frac{\mu}{M} g(\lambda/\omega_c)
\end{eqnarray}
and the propagator (which is proportional to the linear response function) reads
\begin{eqnarray}\label{GL}
\tilde{\mathcal G}_+(\lambda) = \frac{1}{\lambda^2 + \frac{\mu}{M}g(\lambda/\omega_c) + \omega^2 }
\; . 
\end{eqnarray}
Our objective is to find the oscillation frequency and the damping of the impurity motion in the small coupling limit. Thus, we need the 
inverse Laplace transform of Eq.~(\ref{GL}) which can be expressed in terms of the Bromwich integral,
\begin{eqnarray}\label{bromwich}
\mathcal G_+(t) = \frac{1}{2\pi i}\int_{c-i\infty}^{c+i\infty}{\rm d}\lambda \ e^{\lambda t} \ \tilde{\mathcal G}_+(\lambda)\;  ,
\end{eqnarray}
where the real number $c$ is greater than the real part of all poles of $\tilde{\mathcal G}_+(\lambda)$. The integral in Eq.~(\ref{bromwich}) 
can be solved by displacing the complex contour towards the left and by evaluating the encountered residues. Hence, we seek for the 
complex points at which the 
denominator of Eq.~(\ref{GL}) vanishes, that is $z^2 + \mu' g(z) + \omega^2/\omega_c^2 = 0$ with
\begin{align}\label{eq:zL}
z = i\sigma - \Gamma , \;\;\; \mathrm{where}\;\; \frac{\Gamma}{\sigma} \ll 1
\;\; \mbox{and} \;\;
\mu' = \frac{\mu}{M\omega_c^2}
\; .  
\end{align}
Note that $z$, $\sigma$, $\Gamma$ and $\mu'$ are dimensionless 
and $\sigma$ and $\Gamma$ real.
From Eq.~(\ref{bromwich}) it is clear that $\Gamma$ corresponds to the exponential damping (measured in units of $\omega_c$) of the propagator and hence $\Gamma$ 
has to be positive (whereas $\sigma$ is the oscillating frequency measured in units of $\omega_c$ and can be positive or negative). 
Thus, the first singularities 
encountered when displacing the contour towards the left are the imaginary axis for which $g(z)$ is obviously singular when $\re z = 0$. In 
other words, we have to account for the difference 
$g(i\sigma+0^+)-g(i\sigma-0^+)$ when displacing the integration contour past the imaginary axis. We will come 
back to this point in just a few lines.
The integral in the second term of the rhs of Eq.~(\ref{eq:gut}) can be recast as
\begin{eqnarray}
\label{eq:10L}
I(-\Gamma,\sigma) 
&\equiv& 
\int_0^\infty{\rm d}\zeta \ \frac{e^{-\zeta}}{\zeta^2+z^2} 
\nonumber\\
&\simeq&
\frac{1}{|\sigma|}\int_0^\infty{\rm d}\zeta \ \frac{e^{-|\sigma|\zeta}}{\zeta^2-1-2i\Gamma/\sigma} 
\nonumber\\
&\simeq&
\frac{1}{|\sigma|}\left[A(|\sigma|) + \sgn\left[\frac{\Gamma}{\sigma}\right]\frac{i\pi}{2} e^{-|\sigma|}\right] 
\nonumber\\
&&
+ \frac{|\Gamma|}{\sigma^2}\left[\frac{\pi}{2}(1+|\sigma|)e^{-|\sigma|} + i\sgn\left[\frac{\Gamma}{\sigma}\right] B(|\sigma|)\right] \; ,
\nonumber
\end{eqnarray}
where we expanded the denominator of the integrand to first order in $\Gamma/\sigma$ and we defined
\begin{eqnarray*}
A(\sigma) &=& \sinh\sigma\Chi\sigma-\cosh\sigma\Shi\sigma
\; , 
\nonumber\\
B(\sigma) &=& \sigma\left[\cosh\sigma\Chi\sigma-\sinh\sigma\Shi\sigma\right]
\nonumber \\
&&
+\cosh\sigma\Shi\sigma-\sinh\sigma\Chi\sigma 
\; ,
\end{eqnarray*}
where $\Shi$ and $\Chi$ are the hyperbolic sine and cosine integrals, respectively.
The contribution from the crossing of the imaginary axis is now easily obtained: 
$\Delta I \equiv I(0^+,\sigma)-I(-0^+,\sigma)=-i\pi e^{-|\sigma|}/\sigma$.
After adding $-z^4 \Delta I / 2$ to $g(z)$ we thus obtain for $\re z < 0$ 
\begin{align}
g(z) &\simeq -\frac{\sigma^2}{2}+\frac{i\pi\sigma^3}{4}e^{-|\sigma|}-\frac{\sigma^3}{2}A(\sigma)\nonumber \\
&\;\; - \frac{\pi\Gamma\sigma^2}{4}(1+|\sigma|)e^{-|\sigma|} 
-\pi\Gamma\sigma^2 e^{-|\sigma|}
\; ,
\end{align}
where we neglected all real terms of $\mathcal O((\Gamma/\sigma)^2)$ and all imaginary terms of $\mathcal O(\Gamma/\sigma)$.

\begin{figure}[h]
  \begin{center}  
   \includegraphics[width=0.35\textwidth]{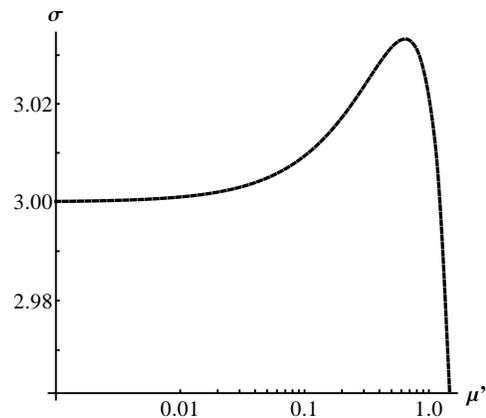} 
  \end{center}
\caption{Dependence of the oscillator frequency $\sigma$ on the impurity-bath coupling $\mu'=\mu/(M\omega_c^2)$ 
for $\omega/\omega_c = 3.0$ obtained from Eq.~(\ref{sigma2}).
A logarithmic scale has been used.
}
\label{fig1}
\end{figure}

\begin{figure}[h]
  \begin{center}  
  \includegraphics[width=0.49\textwidth]{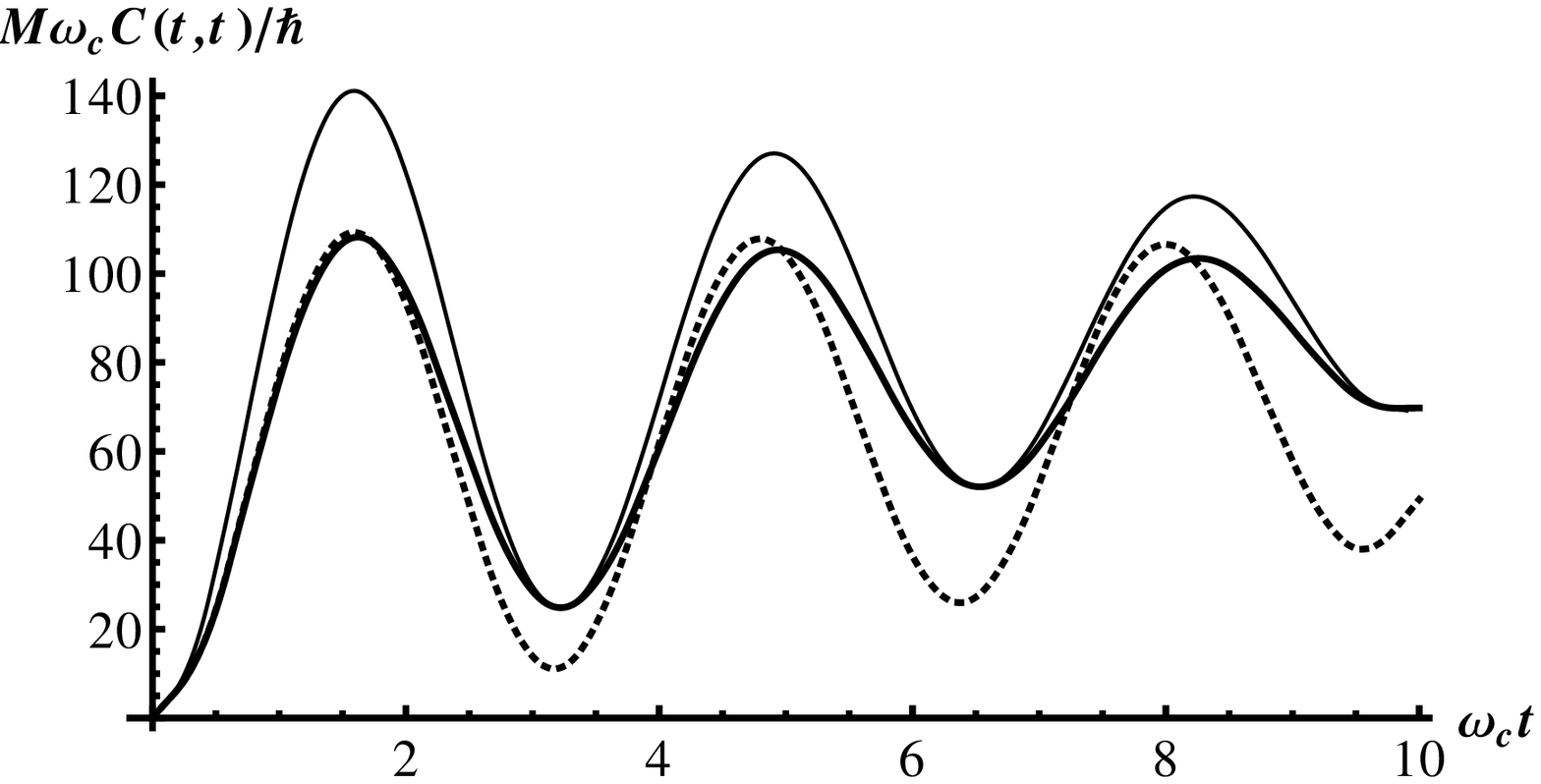}  \\
  \includegraphics[width=0.49\textwidth]{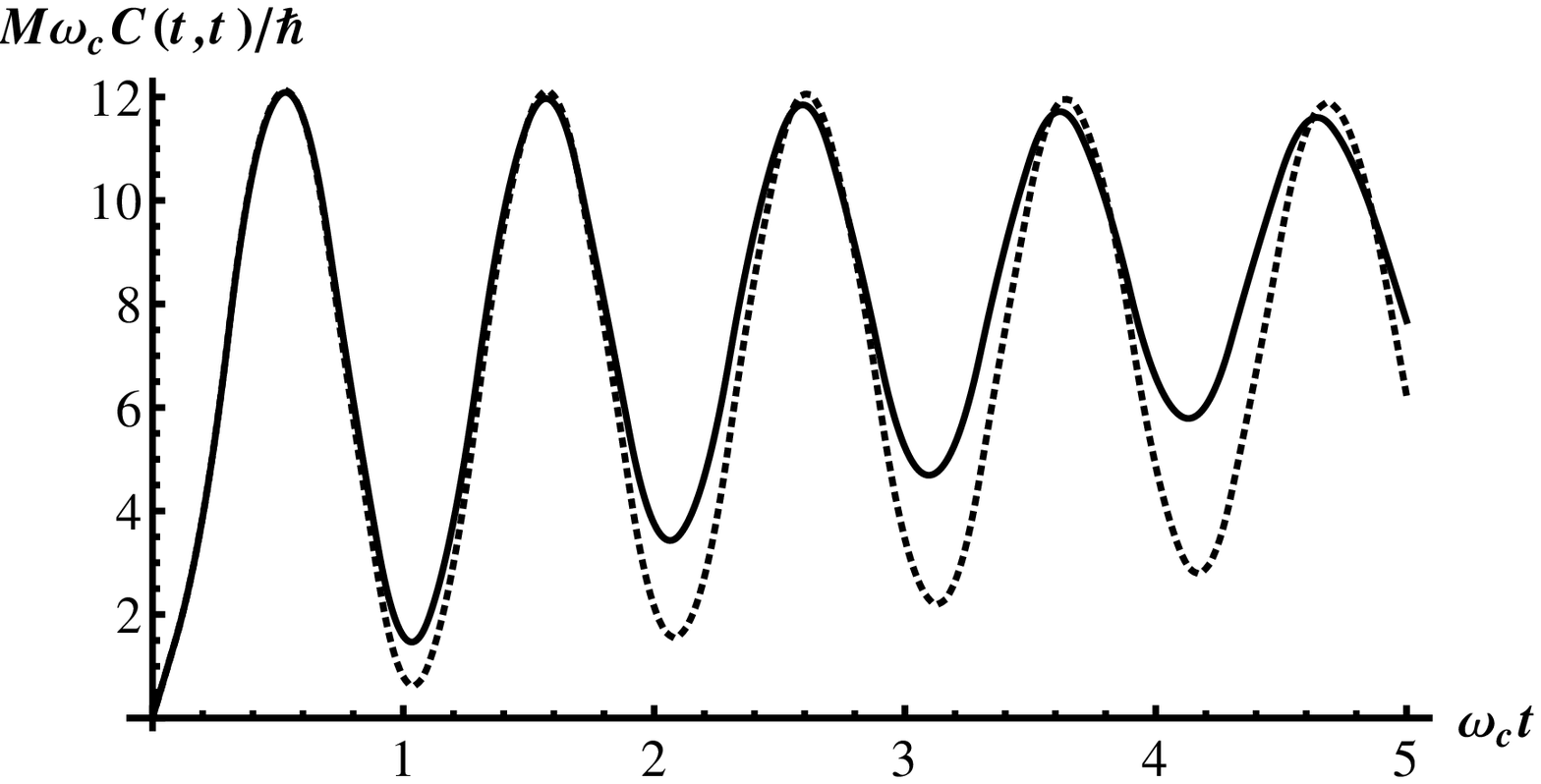}
  \end{center}
\caption{Theoretical results for the variance of the impurity position $C(t,t)$ after its quantum Brownian motion induced by a LL bath in the ``high temperature''--regime which is experimentally available. We chose $1/(4\epsilon^2\omega_c) = 10$ for the thick lines in both images, and $\omega/\omega_c = 1.0$ in the upper and $\omega/\omega_c = 3.0$ in the lower image. The impurity--quantum liquid couplings are $\mu' = 0.5$ (thick straight lines) and $\mu' = 0.2$ (thick dotted lines), respectively. In the upper image the thin line has been obtained with $1/(4\epsilon^2\omega_c) = 50$ and $\mu' = 0.5$.}
\label{fig10}
\end{figure}

At the poles, the real and the imaginary part of the denominator $z^2+\mu'g(z)+(\omega/\omega_c)^2$ vanish simultaneously.
 By using the form of $g(z)$ just derived we see that this can be 
achieved with the choices
\begin{eqnarray}
\label{friction}
&&
\Gamma = \frac{\pi\mu'}{8}\sigma^2 e^{-|\sigma|}
\; , 
\\
\label{sigma2}
&&
\frac{\omega^2}{\omega_c^2} = 
\sigma^2 + \frac{\mu'}{2}\left[\sigma^3 A(\sigma) + \sigma^2\right] + 2\Gamma^2(5+|\sigma|) 
\; .
\end{eqnarray}
Note the symmetry $\sigma \mapsto -\sigma$ in Eq.~(\ref{sigma2}). Equations~(\ref{friction}) and~(\ref{sigma2}) determine the oscillating behavior of $\mathcal G_+(t)$: $\Gamma$ corresponds to the damping and $\sigma$ to the frequency of the oscillation (both quantities measured in units of $\omega_c$).
We assumed the damping $\Gamma$ to be small compared to $\sigma$ which  by virtue of Eq.~(\ref{friction}) translates into
\begin{eqnarray}
\mu' \ll \frac{4}{\pi\sigma}e^{\sigma} \ge \frac{4 e}{\pi} \approx 3.461 
 \; . 
\end{eqnarray}
The behavior of the 
oscillating frequency $\sigma$ (measured in units of $\omega_c$) strongly depends on the trapping frequency $\omega/\omega_c$. Figure~\ref{fig1} shows the dependence of $|\sigma|$ on 
$\mu'=\mu/(M\omega_c^2)$ for the cases $\omega/\omega_c = 1.0$ and $\omega/\omega_c = 3.0$.
When $\hbar\beta\omega \ll 1$ (as in \cite{Giamarchi11}) the properties of the oscillations of $\mathcal G_+(t)$ coincide with the ones for $\mathcal C^{\rm 1eq}(t)$ [see App.~\ref{app:classical}]. Then, by observing Eq.~(\ref{sigma2L}) one deduces that the actual oscillation frequency of $\mathcal C(t,t)$ and its damping are rather $2\sigma$ and $2\Gamma$, respectively.
As can be seen for moderate to high trapping frequencies there is 
an \emph{increase} of the oscillator frequency followed by a decrease 
induced by the bath [see Fig.~\ref{fig1}]. Such a behavior is not observed for small frequencies (compared to $\omega_c$). The experimental results in \cite{Giamarchi11} confirm such a ``peak''
although very few data-points are shown in this paper and the error bars
are quite large so it is hard to draw firm conclusions on the 
actual experimental behavior at this stage. The ``peak'' in the curve of the oscillation frequency can be used in further experiments to determine whether the bath is actually described by a LL or not since it is a direct consequence of the non-Ohmic spectral density (\ref{spectralL}).
%
In Fig.~\ref{fig10} we show the correlator Eq.~(\ref{sigma2L}) for values of the impurity--bath coupling ($\mu' = 0.5$ and $\mu' = 0.2$) in the ``high 
temperature'' regime $\beta\hbar\omega \ll 1$ typical for experiments. The parameters were chosen to be 
$1/(4\epsilon^2\omega_c) = 10$ in both images for the thick lines, $\omega/\omega_c = 1.0$ in the upper and $\omega/\omega_c = 3.0$ in the lower image. The thin line in the upper image has been obtained with $1/(4\epsilon^2\omega_c) = 50$. The curves qualitatively agree with the experimental data \cite{Giamarchi11}. One clearly recognizes the \emph{increase of the oscillation frequency} induced by an increase in the coupling $\mu'$ for $\omega/\omega_c = 3.0$. The typical oscillation width in experiments is about $\rm 15 \mu m$ with a frequency of $\sigma \simeq 550 \rm s^{-1}$. The final position width of a (high temperature) quantum Brownian particle is given by $1/(\beta M\omega^2)$ \cite{Grscin88}. With the mass $M$ of the $^{41}\rm K$ atoms used as an impurity and the bath temperature $T \approx 350 \rm nK$ in \cite{Giamarchi11} this yields $\sqrt{C(t,t)}\approx 15\rm \mu m$ which is even quantitatively the right value.
%
%
%

%

\section{Summary and Outlook}\label{conclude}

Motivated by recent experiments on impurity dynamics in quantum liquids we studied the non equilibrium dynamics of a quantum Brownian particle coupled to a quantum thermal bath of harmonic 
oscillators for generic Gaussian initial conditions. We derived a closed expression for the non equilibrium correlation 
function, which we showed to be easy to derive as variations of a generating functional Eq.~(\ref{eq:Jfinal}). 
We used the analysis in \cite{Grscin88} as a starting point to obtain this generating functional by employing path integral methods. We then 
showed that factorizing initial conditions (where the bath and the particle are initially uncoupled) 
are a special case of the non--factorizing initial conditions on which a position measurement has been performed. We demonstrated 
the correctness of our approach by deriving the equilibrium correlation function without imposing time--translational invariance 
(presented in App.~\ref{sec:FDT}). 

We applied this general formalism to the study of three physical situations. First, we studied the equilibration process of a trapped 
particle after an initial position measurement. In this case we considered Ohmic dissipation. While the classical (high temperature) 
correlator relaxes exponentially on a time scale $\gamma^{-1}$, the low temperature correlator $C(t,t')$ in the strongly
quantum regime shows an algebraic relaxation of the form $1/(tt')^2$ 
which is independent of the dissipation strength $\gamma$. 
Therefore, the information that an initial measurement on the system has been performed persists for a very long time. We then 
showed that the equilibration process is different if, instead of a position measurement, a sudden quench in the trapping potential
is performed at the initial time. We showed that in this case the relaxation 
is exponential in time even in the quantum regime 
with the slight difference that at very low temperatures the relaxation time is of order 
$2\gamma^{-1}$ rather than $\gamma^{-1}$ for high temperatures. Accordingly, the relaxation due to quantum fluctuations is almost as 
effective as thermal relaxation in this case. 

Third, we applied our approach to the motion 
of impurities in 1D ultra cold quantum gases confined in a very elongated harmonic trap. For low excitation energies 
the 1D gas is well described by the Luttinger-theory which predicts a Gaussian theory. We modeled the 
impurity--gas interaction by a simple contact potential. The resulting Hamiltonian is also known as the Fr\"ohlich polaron 
Hamiltonian which features a non-linear 
coupling of the polaron (here the impurity) to the bath (here the Luttinger liquid). It is well known that this Hamiltonian induces a mass shift 
on the impurity depending on the interaction strength. Although the mass shift is clearly observed, no significant 
change of the oscillation frequency is detected in \cite{Giamarchi11}. Hence it is likely that the confining potential is renormalized
as well in such a way that it counteracts the polaronic mass shift. We then proposed to use, 
as a first approximation to this problem, the Gaussian approximation of the polaron Hamiltonian
(more subtle issues linked to polaron and harmonic trap renormalization together with a quantitative comparison between
theory and experiment will be presented elsewhere). The spectral density of the resulting exotic ``bath'' 
turns out to be strongly super-Ohmic with the low frequency spectral density $S(\omega) \sim \omega^3$. With the help of the 
generating functional approach we were able to deduce the non equilibrium correlation function of the impurity 
position. Our formula correctly reproduces the oscillating behavior of the position variance (which is the 
equal time correlation function) as well as the initial momentum of the impurity due to its strong localization at the center of the 
quantum liquid which is a pure quantum effect. We showed that the typical impurity position width measured in \cite{Giamarchi11} quantitatively matches the theoretical value for a quantum Brownian particle. This suggests that non equilibrium quantum Brownian motion theory can describe many aspects of such impurity dynamics. 
The super-Ohmic dissipation leads to a curious phenomenon: upon increasing the coupling 
between the impurity and ``bath'' the oscillation frequency \emph{increases} 
before decreasing. This peculiar behavior is in strong contrast with that induced by Ohmic dissipation, for which the 
damped oscillator has a lower frequency than the free one and it can be used as a signature for Luttinger liquid baths.

To our knowledge our formalism is the only one that can fully access non equilibrium correlation functions in quantum Brownian motion in 
contrast to the density matrix approach in \cite{Grscin88} that yields only one-time quantities. It proved effective for the description of 
impurity dynamics in Luttinger liquids. By using a quantum Langevin equation the authors of \cite{Giamarchi11} have described the 
impurity motion with some success. However, the same authors pointed out that the Langevin approach failed to grasp all aspects of the 
experimental evidence. We are convinced that one can fit our theoretical results to experimental data as soon as more data become 
available in the future. The parameter $\omega_c$ which depends on the actual impurity--quantum liquid coupling can then be 
experimentally determined within our framework. We hope to address this topic in a future study.  

To conclude, our theoretical framework proved effective to deduce all kinds of non equilibrium correlation functions and we obtained 
promising results for its application to impurity motion in 1D systems. Further possible applications of our approach include various topics 
such as the 
study of decoherence problems, the impurity motion in other types of systems or non equilibrium quantum transport phenomena. The 
present work is expected to provide a further 
motivation for a thorough understanding of quantum Brownian motion. 


\appendix
\begin{appendices}

\section{Coherent state path integral formulation}
\label{app1}

In this Appendix we detail the derivation of a path-integral representation of the generating 
functional of the multi-time correlation functions of a quantum particle in contact with a 
generic quantum bath made of an ensemble of harmonic oscillators.

The terms in the Hamiltonian of the dissipative quantum Brownian particle that depend on the bath variables, 
Eqs.~(\ref{eq:Hb}) and~(\ref{eq:Hsb}),  can be rewritten in terms of creation and annihilation operators of the 
bath oscillators in view of a later use of coherent states. One defines the creator and the annihilator 
of the $n$-th oscillator mode by  
\begin{align}\label{eq:aplus}
  \hat a_n^\dagger &= \sqrt{\frac{m_n\omega_n}{2\hbar}}\left[\hat x_n-\frac{i}{m_n\omega_n}\hat p_n\right]
  \; ,\\ 
\label{eq:a}
  \hat a_n &= \sqrt{\frac{m_n\omega_n}{2\hbar}}\left[\hat x_n+\frac{i}{m_n\omega_n}\hat p_n\right] \;.
\end{align}
The operators $\hat a_n$ and $\hat a_n^+$ satisfy the bosonic commutation relations 
$[\hat a_n^+,\hat a_m]=\delta_{n,m}$ and $[\hat a_n^+,\hat a_m^+]=[\hat a_n,\hat a_m]=0$. 
Equations~(\ref{eq:Hb}) and (\ref{eq:Hsb}) read in terms of the $\hat a_n^+$ and $\hat a_n$ 
\begin{align}\label{eq:Hb2}
  \hat{\mathcal H}_B &= \sum_{n=1}^\infty \hbar\omega_n {\hat a}_n^+{\hat a}_n
   \; , \\
\label{eq:Hsb2}
  \hat{\mathcal H}_{SB} &= \sum_{n=1}^\infty g_n {\hat q} ({\hat a}_n^+ + {\hat a}_n) 
  \; .
\end{align}
Here we introduced the notation $g_n \equiv \sqrt{\hbar c_n^2/2m_n\omega_n}$.

We introduce the coherent states of the harmonic oscillators, 
which are particularly suitable when dealing with the bosonic ladder operators in Eq.~(\ref{eq:aplus}), 
\begin{eqnarray}
  |\xi\rangle = e^{\xi\hat a^+}|0\rangle, \qquad\qquad
  \langle\xi| = \langle 0 |e^{\xi^*\hat a}\; ,
\end{eqnarray} 
where $\xi$ is a complex number and $\xi^*$ its complex conjugate. $\hat a^+$ and $\hat a$ stand for the 
creation and annihilation operator of each harmonic oscillator. The coherent states are eigenstates of the 
annihilation operator, that is
\begin{eqnarray}
  \hat a |\xi\rangle = \xi |\xi\rangle \; , 
  \qquad\qquad 
  \langle\xi| \hat a^+ = \langle\xi|\xi^* 
  \; ,
\end{eqnarray}
with the  properties
\begin{eqnarray}
  \langle\xi|\zeta\rangle = e^{\xi^*\zeta} 
  \qquad\mbox{and}\qquad 
  \mathds 1' = \int{\rm d}\xi^*{\rm d}\xi \, e^{-\xi^*\xi}|\xi\rangle\langle\xi|
\; , 
\end{eqnarray}
where $\mathds{1}'$ denotes the unity matrix of one oscillator. Hence, the unity matrix of the whole particle--bath system $\mathds 1$ can be written in terms of the product states $|q,\{\xi_n\}\rangle$ as
\begin{equation}
  \mathds 1 = \int{\rm d} q\int\prod_n\left\{{\rm d}\xi_n^*{\rm d}\xi_n  \, e^{-\xi_n^*\xi_n}\right\}|q,\{\xi_n\}\rangle\langle q,\{\xi_n\}| 
  \; .
\end{equation}
The trace of any observable $\hat B$ that depends on the particle and the bath operators can be expressed as
\begin{equation}\label{eq:trace}
  \mbox{Tr} \hat B = \int{\rm d} q\int\prod_n\left\{{\rm d}\xi_n^*{\rm d}\xi_n \, e^{-\xi_n^*\xi_n}\right\} \langle q,\{\xi_n\}|\hat B |q,\{\xi_n\}\rangle  
  \; .
\end{equation}
The generating functional can now be obtained by supplementing the potential $V(\hat q;s)$ in Eq.~(\ref{eq:Hs}) by a linear term 
$-H(s)\hat q$ where $H(s)$ is a $c$-number function that plays the role of an external source. To be more explicit, we introduce two 
distinct sources $H(s)$ and $H'(s)$ in the potential $V$ for the time evolution operator
\begin{align}
\mathcal K(q_f,\{\xi_{n,f}\};&q_i,\{\xi_{n,i}\};t) \equiv \nonumber\\
&\langle q_f,\{\xi_{n,f}\}|\hat{\mathcal T} e^{-\frac{i}{\hbar}\int_0^{t} \!\! dt' \ \mathcal H(t')} 
  |q_i,\{\xi_{n,i}\}
  \rangle
\end{align}
and its Hermitian conjugate
\begin{align}
  \mathcal K^*(q'_f,\{\xi'_{n,f}\};&q'_i,\{\xi'_{n,i}\};t) \equiv \nonumber\\
&\langle q'_i,\{\xi'_{n,i}\}|\hat{\mathcal T}^\dagger 
  e^{\frac{i}{\hbar} \int_0^t \!\! dt' \mathcal H(t')} 
  |q'_f,\{\xi'_{n,f}\}\rangle \; , 
\end{align}
respectively. 
$\hat{\mathcal T}^\dagger$ is the anti-chronological time ordering operator. In the following we will use the shorthand notation
$
{\rm d}\xi_i = \prod_{n=1}^\infty\ e^{-\xi_{n,i}^*\xi_{n,i}}{\rm d}\xi_{n,i}^*{\rm d}\xi_{n,i} \ .
$ 
All correlation functions of the position $\hat q$ can be obtained by taking the corresponding variations of the trace of the 
\emph{time-dependent} density matrix $\hat \rho(t) \equiv \hat{\mathcal K}\hat\rho_0\hat{\mathcal K}^*$, 
Eq.~(\ref{eq:densityT}), 
with respect to $H(s)$ and $H'(s)$. The matrix elements of  
$\hat \rho(t) \equiv \hat{\mathcal K}\hat\rho_0\hat{\mathcal K}^*$ are given by 
\begin{align}
&
\mathcal W(q_f,\{\xi_{n,f}\};q'_f,\{\xi'_{n,f}\};t) = \langle q_f,\{\xi_{n,f}\}|\ \hat\rho(t)\ |q'_f,\{\xi'_{n,f}\}\rangle
\; ,
\\
&
\mathcal W(q_i,\{\xi_{n,i}\};q_i',\{\xi_{n,i}'\}) = \langle q_i,\{\xi_{n,i}\}|\ \hat\rho_0 \ |q_i',\{\xi_{n,i}'\}\rangle 
\; .
\end{align}
The path integral representation of $\mathcal K$ and $\mathcal K^*$ are found by using standard methods:
\begin{align}\label{eq:Kapp}
 \mathcal K(q_f,\{\xi_{n,f}\};q_i,\{\xi_{n,i}\};t) &= 
  \int\mathcal D q^+\mathcal D \xi^+  \nonumber\\
&\times\exp\left(\frac{i}{\hbar}\mathcal S[q^+,\{\xi^+_n\}]\right) 
 \\ 
\label{eq:Kappstar} 
 \mathcal K^*(q'_f,\{\xi'_{n,f}\};q'_i,\{\xi'_{n,i}\};t') &= 
 \int\mathcal D q^- \mathcal D \xi^- \nonumber\\
&\times \exp\left(-\frac{i}{\hbar}\mathcal S^*[q^-,\{\xi^-_n\}]\right)\; ,
\end{align} 
where we make clear by the superscripts $^+$ and $^-$ which path belongs to $\mathcal K$ and which to $\mathcal K^*$, respectively. 
The real time interval $[0,t]$ has been discretized into $T \in \mathds{N}$ steps of length $\Delta t$ with $t=\Delta t T$. 
The  functional integration measures are defined as $\mathcal D q = \prod_{j=1}^{T-1} {\rm d} q_j$
with $q_j \equiv q(jt/T)$.
The terms contributing to the total action, $\mathcal S[q,\xi_n] = \mathcal S_S[q] + \mathcal S_B[\xi_n] + \mathcal S_{SB}[q,\xi_n]$ 
introduced in Eqs.~(\ref{eq:Kapp}) and~(\ref{eq:Kappstar}), read in discretized form 
\begin{align}\label{eq:action1}
\mathcal S_S[q] &= \sum_{j=1}^T \Delta t\left[\frac{M}{2}\left(\frac{q_j-q_{j-1}}{\Delta t}\right)^2 - V(q_j;j\Delta t)  \right.\nonumber\\
&\left. + H(j\Delta t) q_j \right] \; , 
\\
\mathcal S_B[\xi_n] &+ \mathcal S_{SB}[q,\xi_n] = i\hbar \sum_{j=1}^{T-1}\xi^*_{n,j}(\xi_{n,j}-\xi_{n,j-1}) 
\nonumber\\
&+ \Delta t \sum_{j=1}^T\left[\hbar\omega\xi^*_{n,j}\xi_{n,j-1} + g_n q_j (\xi^*_{n,j}+\xi_{n,j-1}) \right] 
\; . 
\end{align}

The  reduced  density matrix  depends only on the particle variables and the external sources,  
\begin{align}\label{eq:reddens}
&\mathcal W(q_f;q_f';t) \equiv \int{\rm d} q_i{\rm d} q'_i{\rm d}\xi_i{\rm d}\xi_i'{\rm d}\xi_f
\
\mathcal K(q_f,\{\xi_{n,f}\};q_i,\{\xi_{n,i}\};t)
\nonumber\\
&
\times \mathcal W(q_i,\{\xi_{n,i}\};q'_i,\{\xi'_{n,i}\})\mathcal K^*(q'_f,\{\xi_{n,f}\};q'_i,\{\xi'_{n,i}\};t)
\; . 
\end{align}
In Eqs.~(\ref{eq:Kapp}) and~(\ref{eq:Kappstar}) we omitted normalization factors that do not depend on the bath nor on the particle 
variables. 
Note that the integral runs only over bath and particle variables with an index between $1$ and $T-1$ since 
$q_0 = q_i$, $q_T = q_f$ (and analogously for the bath variables) are fixed for $\mathcal K$ and $q_0 = q'_f$, $q_T = q'_i$ (and 
analogously for the bath variables) are fixed for $\mathcal K^*$.

The path integral description of $\hat \rho_0$ is obtained by dividing the imaginary time interval $[0,\beta\hbar]$ into $T$ time steps. 
Consequently, by using $\hat\rho=e^{-\beta \hat{ \mathcal H}_0}$ we find
\begin{align}\label{eq:W}
\mathcal W(q_i',\{\xi_{n,i}'\};&q_i,\{\xi_{n,i}\}) = \nonumber\\
&\int\mathcal D q^0\prod_n\left\{ \mathcal D \xi^0_n \ \exp\left(-\frac{1}{\hbar}\mathcal S_0[q^0,\xi^0_n]\right)\right\} \; , 
\end{align} 

where $\mathcal S_0[q,\xi_n] = \mathcal S_{0S}[q] + \mathcal S_{0B}[\xi_n] + \mathcal S_{0SB}[q,\xi_n]$ with
\begin{align}\label{eq:action2}
\mathcal S_{0B}[\xi_n] &+ \mathcal S_{0SB}[q,\xi_n] 
= 
\hbar \sum_{j=1}^{T-1}\xi^*_{n,j}(\xi_{n,j}-\xi_{n,j-1}) 
\nonumber\\
&+ \Delta t' \sum_{j=1}^T\left[\hbar\omega\xi^*_{n,j}\xi_{n,j-1} + g_n q_j (\xi^*_{n,j}+\xi_{n,j-1}) \right]\; .
\end{align}
We introduced the imaginary time path step $\Delta t' = \beta\hbar / T$. 

\subsection{Integration over the bath variables}\label{app2}

The influence functional in discretized form reads
\begin{align}\label{eq:F1}
&
\mathcal F[\{q_j\}] 
= \nonumber\\
&\prod_n \int\prod_{j=1}^{3T}{\rm d}\xi_{n,j}{\rm d}\xi^*_{n,j} \ e^{-\sum_{j,j'=1}^{3T}\xi_{n,j}^*K^{-1}(j,j')\xi_{n,j'}} \nonumber\\
&e^{\frac{i}{\hbar}g_n\sum_{j=1}^T\Delta t q_j(\xi_{n,j} + \xi_{n,j}^*)}
\nonumber\\
&\times e^{ -\frac{1}{\hbar} g_n\sum_{j=T+1}^{2T}\Delta t' q_j(\xi_{n,j}+\xi_{n,j}^*) - \frac{i}{\hbar} g_n\sum_{j=2T+1}^{3T}\Delta t q_j(\xi_{n,j}+\xi_{n,j}^*)} \; ,
\end{align}
which depends on the correlation matrix 
\begin{equation}\label{eq:Kinverse}
  K^{-1} = \left(
  \begin{array}{cccccc}
    1 & 0 & 0 & \cdots & & -k_1 \\
    -k_2 & 1 & 0 & \cdots & & 0 \\
    0 & -k_3 & 1 & \cdots & & 0 \\
    0 & 0 & -k_4 & 1 & \cdots & 0 \\
    \vdots \\
    0 & 0 & \cdots & & -k_{3T} & 1
  \end{array}
  \right) \; ,
\end{equation}
with $k_j = 1 + i\Delta t \omega$ for $j \le T$, $k_j = 1 - i\Delta t \omega$ for $j \ge 2T$ and 
$k_j = 1 - \Delta t' \omega$ for $T< j \le 2T$. Note that in this notation there are $3T$ time steps in Eq.~(\ref{eq:F1}) and $q_j = q_j^+$ for $j \le T$, $q_j = q_j^0$ for $T<j\le 2T$ and $q_j = q_j^-$ for $j > 2T$ is understood with an analogous notation for the $\xi_{n,j}$. 

The part of the action that depends only on the particle variables $\mathcal S_S$ can be found by combining the relevant contributions in Eqs.~(\ref{eq:Kapp}),~(\ref{eq:Kappstar}) and (\ref{eq:W}),
\begin{align}
\frac{i}{\hbar}\tilde{\mathcal S}_S(\{q_j\}) = \frac{i}{\hbar} \mathcal S_S(\{q_j\}_{j=1}^T) &- \frac{i}{\hbar} \mathcal S_S(\{q_j\}_{j=2T+1}^{3T}) \nonumber\\
&- \frac{i}{\hbar} \mathcal S_{0S}(\{q_j\}_{j=T+1}^{2T}) \; .
\end{align}
Note that the exponential factors that stem from the bath integration measure Eq.~(\ref{eq:dxi}) exactly combine with the sums in the actions (\ref{eq:action1}) and (\ref{eq:action2}). The elements $K(j,j')$ of the matrix $K$ are easily found:
\begin{align}
\label{eq:Kjj}
&K(j,j') = \frac{1}{1-k_1\cdots k_{3T}}  
   \left\{
  \begin{array}{ll}
    1 & \mbox{for} \;\;  j=j' \\
    \prod_{l=j'+1}^j k_l & \mbox{for} \;\;  j>j' \\
    \frac{k_1\cdots k_{3T}}{\prod_{l=j+1}^{j'} k_l} & \mbox{for} \;\; j<j' \; .
\end{array} 
\right.
\end{align}
The Gaussian integral in Eq.~(\ref{eq:F1}) is now readily done. Explicit expressions of Eq.~(\ref{eq:Kjj}) in the continuum limit $T \to \infty$ are easily obtained: for instance, when $j,j'<T$, $K(j,j')$ couples to two $q^+$ fields and is given by
\begin{align}
&K(j,j') = \frac{1}{1-e^{-\beta\hbar\omega}}
\left\{
  \begin{array}{ll}
    e^{i\omega\Delta t (j-j')} & \mbox{for} \;\;  j'<j \le T \\
    e^{-\beta\hbar\omega + i\omega\Delta t (j-j')} & \mbox{for} \;\;  j<j' \le T \; .
  \end{array}
  \right.
\end{align}
Note that under the sum over $j$ and $j'$ only its symmetrized version occurs.
By reintroducing the fields $q^+$, $ q^-$ and $q^0$ we find~\cite{Grscin88}
\begin{eqnarray}
\mathcal F[q^+,q^-,q^0] = \exp\left(-\frac{1}{\hbar}\Phi[q^+,q^-,q^0]\right)\; ,
\end{eqnarray}
where the exponent reads 
\begin{align} \label{eq:phi1}
&\Phi[q^+,q^-,q^0] =  \nonumber\\
&-\int_0^{\beta\hbar}{\rm d}\tau\int_0^\tau{\rm d}\sigma \ K(-i\tau+i\sigma)q^0(\tau)q^0(\sigma) 
  + \int_0^{\beta\hbar}{\rm d}\tau \frac{\mu}{2} {q^0}^2(\tau) \nonumber\\
&- i\int_0^{\beta\hbar}{\rm d}\tau\int_0^t{\rm d} s \ K^*(s-i\tau)q^0(\tau)\left[q^+(s)-q^-(s)\right] \\
&+ \int_0^t{\rm d} t\int_0^s{\rm d} u \left[q^+(s)-q^-(s)\right]\left[K(s-u)q^+(u)\right.\nonumber\\
&\left.-K^*(s-u)q^-(u)\right] \nonumber\\
&+ i\int_0^t{\rm d} s \frac{\mu}{2}\left[{q^+}^2(s)-{q^-}^2(s)\right] \; . \nonumber
\end{align}
The kernel $K$ reads for complex times $\theta = s-i\tau$, $0 \le \tau \le \beta\hbar$
\begin{eqnarray}\label{app:K}
  K(\theta) = \sum_n^\infty\frac{{g_n}^2}{\hbar} \frac{\cosh[\omega_n(\beta\hbar/2-i\theta)]}{\sinh[\omega_n\beta\hbar/2]} \;,
\end{eqnarray}
and the constant $\mu$ is given by 
\begin{eqnarray}
  \mu = 2\sum_n^\infty\frac{{g_n}^2}{\hbar\omega_n} \; .
\end{eqnarray}
Note that for a fermionic bath the only difference lies in the boundary conditions enforced by the trace operation: for fermions anti-periodic boundary conditions apply in contrast to periodic boundary conditions for bosons. This difference is incorporated by replacing $-k_1$ by $k_1$ in Eq.~(\ref{eq:Kinverse}). The analysis for bosons can then be repeated, leading to a fermionic bath kernel where the $\cosh$ and the $\sinh$ in Eq.~(\ref{app:K}) interchange their positions.

The environment can be regarded as a proper heat bath only if the spectrum of the harmonic oscillators becomes quasi-continuous. 
Accordingly, we introduce the spectral density of the bath through
\begin{eqnarray}
  S(\omega) = \pi\sum_n\frac{{g_n}^2}{\hbar}\delta(\omega - \omega_n) = \pi\sum_n\frac{c_n^2}{2m_n\omega_n}\delta(\omega-\omega_n) \; .
\end{eqnarray}
Then the kernel $K(\theta)$ and the constant $\mu$ are rewritten in terms of the spectral density
\begin{eqnarray}\label{eq:kernel}
K(\theta) = \int_0^\infty\frac{{\rm d}\omega}{\pi}S(\omega) \frac{\cosh[\omega(\beta\hbar/2-i\theta)]}{\sinh[\omega\beta\hbar/2]}
\end{eqnarray}
and
\begin{eqnarray}
\mu = 2\int_0^\infty\frac{{\rm d}\omega}{\pi}\frac{S(\omega)}{\omega} \; .
\end{eqnarray}
The real and the imaginary parts of the kernel $K(\theta) = K_R(\theta) + iK_I(\theta)$ are found to be
\begin{align} \label{app:KR}
&
K_R(s-i\tau) = 
\int_0^\infty\frac{{\rm d}\omega}{\pi}S(\omega) \frac{\cosh[\omega(\beta\hbar/2-\tau)]}{\sinh[\omega\beta\hbar/2]}\cos(\omega s)
\\
&
 \label{app:KI}
  K_I(s-i\tau) = 
  -\int_0^\infty\frac{{\rm d}\omega}{\pi}S(\omega) \frac{\sinh[\omega(\beta\hbar/2-\tau)]}{\sinh[\omega\beta\hbar/2]}\sin(\omega s) \; .
\end{align}
The imaginary time argument $\tau$ varies in the interval $[0,\beta\hbar]$ so that it is convenient to introduce the Fourier series of $K(s-i\tau)$ with respect to $\tau$. Introducing the Matsubara frequencies
\begin{eqnarray}
\nu_k = \frac{2\pi k}{\beta\hbar}
\end{eqnarray}
we find 
\begin{eqnarray}\label{eq:KR}
K_R(s-i\tau) = \frac{1}{\beta\hbar} \sum_{k=-\infty}^\infty\ g_k(s) e^{i\nu_k\tau}
\end{eqnarray}
and
\begin{eqnarray}\label{eq:KI}
K_I(s-i\tau) = \frac{i}{\beta\hbar} \sum_{k=-\infty}^\infty\ f_k(s) e^{i\nu_k\tau}
\; ,
\end{eqnarray}
where the functions $g_k$ and $f_k$ are defined through
\begin{eqnarray}\label{app:gk}
g_k(s) = \int_0^\infty\frac{{\rm d}\omega}{\pi} S(\omega)\frac{2\omega}{\omega^2 + \nu_k^2}\cos(\omega s)
\end{eqnarray}
and 
\begin{eqnarray}\label{app:fk}
f_k(s) = \int_0^\infty\frac{{\rm d}\omega}{\pi} S(\omega)\frac{2\nu_k}{\omega^2 + \nu_k^2}\sin(\omega s) 
\; .
\end{eqnarray}
In the following we will express most quantities in terms of the functions $g_k$ and $f_k$. For real times 
the real and the imaginary parts of the kernel (\ref{eq:kernel}) read [see Eqs.~(\ref{eq:KR}) and (\ref{eq:KI})] 
\begin{eqnarray}
K_R(s) = \int_0^\infty \frac{{\rm d}\omega}{\pi} S(\omega) \coth(\beta\hbar\omega/2)\cos(\omega s)
\end{eqnarray}
and
\begin{eqnarray}
K_I(s) = -\int_0^\infty \frac{{\rm d}\omega}{\pi} S(\omega) \sin(\omega s) \; .
\end{eqnarray}
We now eliminate the local terms in Eq.~(\ref{eq:phi1}). We define the two new kernels
\begin{eqnarray}\label{eq:gamma}
\gamma(s) = \frac{2}{M}\int_0^\infty\frac{{\rm d}\omega}{\pi} \frac{S(\omega)}{\omega} \cos(\omega s)
\end{eqnarray}
and
\begin{eqnarray}\label{eq:smallk}
k(\tau) = \frac{M_0}{\beta\hbar} \sum_{k=-\infty}^\infty \zeta_k e^{i\nu_k\tau} \; ,
\end{eqnarray}
where $\zeta_k$ is defined by
\begin{align}\label{eq:zeta}
\zeta_k &= \frac{1}{M_0}[\mu - g_k(0)] \nonumber\\
&= \frac{1}{M_0}\int_0^\infty\frac{{\rm d}\omega}{\pi}\frac{S(\omega)}{\omega}\frac{2\nu_k^2}{\omega^2 + \nu_k^2} \; .
\end{align}
The latter kernel is related to $K_R(-i\tau)$ via 
\begin{align}\label{eq:kKR}
&-\int_0^{\hbar\beta}{\rm d}\tau\int_0^\tau{\rm d}\sigma K_R(-i\tau+i\sigma)f(\tau,\sigma) = \\
&-\frac{\mu}{2}\int_0^{\hbar\beta}{\rm d}\tau f(\tau,\tau) + \frac{1}{2}\int_0^{\hbar\beta}{\rm d}\tau{\rm d}\sigma k(\tau-\sigma)f(\tau,\sigma) \; ,
\nonumber 
\end{align}
with a generic function $f$. 
In terms of the kernels $\gamma(s)$, $k(\tau)$, $K^*(s-i\tau)$, $K_R(s-u)$ and the linear combinations
\begin{eqnarray}\label{eq:xxbar}
  x = (q^+ + q^-)/2 \;\;\;\;\;\mathrm{and}\;\;\;\;\; \bar x = q^+ - q^-
\end{eqnarray}
the exponent of the influence functional reads
\begin{eqnarray}\label{app:phi2}
&&
\Phi[x,\bar x,q^0] = 
\frac{1}{2} \int_0^{\beta\hbar}\!\!\!\! {\rm d}\tau{\rm d}\sigma \ k(\tau-\sigma)q^0(\tau)q^0(\sigma) \nonumber\\
&&
\qquad
- i\int_0^{\beta\hbar} \!\!\!\! {\rm d}\tau\int_0^t \!\!\! {\rm d} s \ K^*(s-i\tau)q^0(\tau)\bar x(s) \nonumber\\
&&
\qquad
+ \frac{1}{2}\int_0^t \!\! {\rm d} s{\rm d} u \ K_R(s-u)\bar x(s) \bar x(u) \\
&&
\qquad 
+i M \int_0^t \!\! {\rm d} s \ \bar x(s)\frac{{\rm d}}{{\rm d} s}\int_0^s{\rm d} u \ \gamma(s-u)\dot x(u)
\; .
\nonumber
\end{eqnarray}
Details of the derivation of Eq.~(\ref{app:phi2}) can be found in the thorough analysis in 
\cite{Grscin88}.


\section{Classical Brownian particle in a harmonic potential: Initial position measurement and quenches in the trapping potential}\label{app:classical}

This part is meant to be a reminder on classical stochastic motion induced by generic baths. 
None of the results presented herein are new but they are useful to be confronted with the quantum 
results discussed in the body of the paper.

The classical Brownian motion of a particle confined in a harmonic potential can be 
described by the \emph{Langevin equation}
\begin{eqnarray}\label{langevin}
\ddot{q}(t) + \int_0^t{\rm d} s \ \gamma(t-s)\dot q(s) + \omega^2 q(t) = \xi(t) 
\; ,
\end{eqnarray}
where $\xi$ is a zero mean Gaussian noise 
\footnote{The underlying probability distribution is of the (Gaussian) Boltzmann-Gibbs type $\exp(-\beta\mathcal H)$ with $\mathcal H$ the \emph{full coupled} Hamiltonian of the particle--bath system. Equation~(\ref{langevin}) thus describes the case where the harmonic oscillator bath and the particle are initially \emph{coupled} as in the quantum case studied in the present work. This subtle point is often overlooked. For more details see p.~21-23 in \cite{We08}.}
with correlation $\langle \xi(t)\xi(s) \rangle = 
\frac{1}{M\beta}\gamma(|t-s|)$ and with $\gamma(t)$ given in Eq.~(\ref{gamma}) \cite{We08}. In the Laplace transform 
formulation, the solution to Eq.~(\ref{langevin}) reads 
\begin{eqnarray}
\tilde q(\lambda) = \tilde{\mathcal G}_+(\lambda) \left[\tilde \xi(\lambda) + v_0 + (\lambda+\tilde\gamma(\lambda))q^0\right] \; ,
\end{eqnarray} 
where we used $\tilde{\mathcal G}_+$ defined in Eq.~(\ref{eq:Gplus}) and
we introduced the initial conditions $q(0) = q^0$ and $\dot q(0) = v_0$. The correlation function is now easily computed and 
it reads
\begin{align}\label{app:correlc}
\tilde{\mathcal C}(\lambda,\kappa) 
&= 
\langle \tilde q(\lambda) \tilde q(\kappa)\rangle = \frac{1}{\beta M}\frac{\tilde\gamma(\lambda)+
\tilde\gamma(\kappa)}{\lambda + \kappa}\tilde{\mathcal G}_+(\lambda)\tilde{\mathcal G}_+(\kappa) \nonumber\\
&+ \tilde{\mathcal G}_+(\lambda)\tilde{\mathcal G}_+(\kappa)\left[v_0^2 + 
(\lambda+\tilde\gamma(\lambda))(\kappa+\tilde\gamma(\kappa)) {q^0}^2 \right] \nonumber\\
&+ \tilde{\mathcal G}_+(\lambda)\tilde{\mathcal G}_+(\kappa) \ v_0 q^0 \left[\lambda+\tilde\gamma(\lambda)
+ \kappa+\tilde\gamma(\kappa) \right] \; ,
\end{align}
where we used the fact that the Laplace transform of $\gamma(|t-s|)$ with respect to 
$t$ and $s$ is given by $[\tilde\gamma(\lambda)+\tilde\gamma(\kappa)]/(\lambda+\kappa)$. 
The initial values $q^0$ and $v_0$ can be sharp or drawn from a probability distribution which is typically of the Maxwell-Boltzmann type, 
that is
\begin{align}\label{Pv0q0}
&P[q^0,v_0] = \frac{\beta M_0 \omega_0}{2\pi} \exp\left[-\beta\left(\frac{M_0}{2}v_0^2 + \frac{M_0}{2}\omega_0^2 {q^0}^2\right)\right] 
\; ,
\end{align}
where $\omega_0$ is the frequency of the initial trapping potential and $M_0$ is the initial mass. From Eq.~(\ref{Pv0q0}) we easily derive
\begin{eqnarray}
\langle {q^0}^2 \rangle  = (\beta M_0 \omega_0^2)^{-1} 
\;\;\;\mathrm{and}\;\;\;\langle v_0^2 \rangle  = (\beta M_0)^{-1}
\; .
\end{eqnarray}
As long as $\omega_0 = \omega$ and $M = M_0$ the correlation function can be rewritten as
\begin{align}\label{app:corrclassical}
\tilde{\mathcal C}^{\rm eq}(\lambda,\kappa) &= \frac{\tilde{\mathcal C}^{\rm 1eq}(\lambda)
+ \tilde{\mathcal C}^{\rm 1eq}(\kappa)}{\lambda+\kappa} \;\;\;\;\;\mathrm{with}
\nonumber \\
\mathcal C^{\rm 1eq}(\lambda) &\equiv \frac{1}{\beta M \omega^2} \frac{\tilde\gamma(\lambda) + \lambda}{\lambda^2 + 
\tilde\gamma(\lambda) + \omega^2} \; ,
\end{align}
which is  the equilibrium correlation function. 

The non equilibrium correlation can be recast in the form
\begin{align}\label{app:corrclassical2}
\tilde C(\lambda,\kappa) &=
\frac{\tilde{\mathcal C}^{\rm 1eq}(\lambda)
+ \tilde{\mathcal C}^{\rm 1eq}(\kappa)}{\lambda+\kappa}
+ \tilde{\mathcal G}_+(\lambda)\tilde{\mathcal G}_+(\kappa)\left[v_0^2 -\frac{1}{\beta M}\right] \nonumber \\
&+ \beta M\omega^2 \ \tilde{\mathcal C}^{\rm 1eq}(\lambda)\tilde{\mathcal C}^{\rm 1eq}(\kappa)\left[\beta M\omega^2 {q^0}^2 - 1\right] \nonumber\\
&+ \tilde{\mathcal G}_+(\lambda)\tilde{\mathcal G}_+(\kappa) \ v_0 q^0 \left[\lambda+\tilde\gamma(\lambda)
+ \kappa+\tilde\gamma(\kappa) \right] \; .
\end{align}
In many cases $q^0$ and $v_0$ are uncorrelated random variables. Then Eq.~(\ref{app:corrclassical2}) transforms into
\begin{align}\label{app:corrclassical3}
\tilde {\mathcal C}(\lambda,\kappa) &= \frac{\tilde{\mathcal C}^{\rm 1eq}(\lambda)
+ \tilde{\mathcal C}^{\rm 1eq}(\kappa)}{\lambda+\kappa}
+ \tilde{\mathcal G}_+(\lambda)\tilde{\mathcal G}_+(\kappa)\left[\langle v_0^2 \rangle -\frac{1}{\beta M}\right] \nonumber \\
&+ \beta M\omega^2 \ \tilde{\mathcal C}^{\rm 1eq}(\lambda)\tilde{\mathcal C}^{\rm 1eq}(\kappa)\left[\beta M\omega^2 \langle{q^0}^2\rangle - 1\right]  \; .
\end{align}
%


\section{The equilibrium initial condition}
\label{sec:FDT}

In this Appendix we use Eq.~(\ref{eq:Jfinal}) in the particular case of an equilibrium initial condition and a 
subsequent evolution still in equilibrium. We show how to derive the equilibrium correlation function 
and we prove that the fluctuation-dissipation theorem (FDT) is satisfied without imposing time-translational invariance 
(TTI)  as has been done before in the literature~\cite{Grscin88}. 


\subsection{The fluctuation-dissipation theorem}

The linear response is easily found by noting that the external source $F(s)$ represents a physical drift force. 
Therefore, by calculating
\begin{align}
\langle \hat q(t) \rangle = \frac{\hbar}{i}\frac{\delta}{\delta G(t)} \exp\left(\mathcal J[F,G]\right)|_{G\equiv 0}
= \int_0^t{\rm d} s \ \mathcal R(t-s) F(s) 
\end{align}
for $F \neq 0$ one finds the response function $\mathcal R(t)$ with respect to the external force $F(t)$. We set $\epsilon \to \infty$ which 
corresponds to the absence of any initial measurement. By using Eq.~(\ref{eq:Jfinal}) we obtain
\begin{align}\label{eq:chi}
\mathcal R(t) = \frac{1}{M} \mathcal G_+(t) 
\qquad \mbox{and} \qquad 
\tilde {\mathcal R}(\lambda) = \frac{1}{M}\frac{1}{\omega^2 + \lambda\tilde\gamma(\lambda) + \lambda^2} 
\; ,
\end{align}
in the time and Laplace transform domains, respectively. These expressions are 
independent of the initial condition. Therefore, the response function is the same in and out of equilibrium. 
Moreover, it is equal to the response function of a classical Brownian particle \cite{Bocuga11} if 
it is coupled to a bath with the same friction kernel $\gamma$.

We will confirm the validity of the FDT when the system is in equilibrium. We choose the initial Hamiltonian to be 
equal to the ``bulk'' one, that is 
\begin{eqnarray}
\omega = \omega_0\;\;\;\mathrm{and}\;\;\;M = M_0 
\; ,
\end{eqnarray}
so that the initial density matrix is equal to the Boltzmann weight $\exp(-\beta\mathcal H)$ with the terms contributing to  
$\mathcal H$ given in Eqs.~(\ref{eq:Hs}),~(\ref{eq:Hb}) and~(\ref{eq:Hsb}). The initial state is not perturbed by any measurement, so 
we take $\epsilon \to \infty$ which implies $\Lambda' = \Lambda$ and $\Omega' = \Omega$. From Eq.~(\ref{eq:corrgen}) we find 
the equilibrium correlation function $\mathcal C^{\rm eq}(t,t')$ which in the Laplace transform version reads
\begin{align}\label{eq:corr2}
&\tilde {\mathcal C}^{\rm eq}(\lambda,\kappa) = \frac{\hbar}{M}\tilde{\mathcal G}_+(\lambda)\tilde{\mathcal G}_+(\kappa)
  \left\{
  \Lambda\lambda\kappa + \frac{\Lambda}{M} \lambda \tilde C_1(\kappa) 
  \right. 
  \nonumber\\
&
\left. 
+ \frac{\Lambda}{M} \kappa \tilde C_1(\lambda) + \Omega 
- \frac{1}{M}\tilde C_2(\lambda) - \frac{1}{M}\tilde C_2(\kappa) + \frac{1}{M^2}\tilde{R''}(\lambda,\kappa) \right\} 
 \; .
\end{align}
This expression can be greatly simplified. We first note that from the definitions of $g_k$ and $f_k$ in Eqs.~(\ref{app:gk}) 
and~(\ref{app:fk}) it follows that
\begin{eqnarray}
\dot{f}_k(s) = \nu_k g_k(s)
\; ,
\qquad 
\tilde{f}_k(\lambda) = \frac{\nu_k}{\lambda}\tilde g_k(\lambda) 
\; ,
\end{eqnarray}
where we used $f_k(0) = 0$. 
The Laplace transform of the kernel 
\begin{align}
R''(s,s') &= \frac{1}{\beta\hbar}\sum_k u_k[g_k(s)g_k(s') \nonumber\\
&- f_k(s)f_k(s')] + M_0 K_R(s-s')
\end{align}
 [see Eq.~(\ref{eq:def10}) and~(\ref{Rtilde}) for $\Lambda = \Lambda'$] can now be written as
\begin{align}
\tilde{R''}(\lambda,\kappa) &= \frac{1}{\beta\hbar}\sum_k u_k\left(1-\frac{\nu_k^2}{\lambda\kappa}\right)\tilde g_k(\lambda) \tilde g_k(\kappa) \nonumber\\
&+ \frac{M}{\beta\hbar}\sum_k \frac{\tilde g_k(\lambda)+\tilde g_k(\kappa)}{\lambda+\kappa}
\end{align} 
and by defining $\tilde h_k(\lambda) = \tilde g_k(\lambda)/M + \lambda$ 
we find that the expression in the curly brackets in the rhs of Eq.~(\ref{eq:corr2}) can be recast as
\begin{align}\label{eq:123}
\frac{1}{\beta\hbar}\sum_k u_k\left(1-\frac{\nu_k^2}{\lambda\kappa}\right)\tilde h_k(\lambda) \tilde h_k(\kappa) 
+ \frac{1}{\beta\hbar}\sum_k \frac{\tilde h_k(\lambda) + \tilde h_k(\kappa)}{\lambda+\kappa} 
\; ,
\end{align}
where we used Eqs.~(\ref{C1C2}) and~(\ref{eq:def1}). 
By combining the expression for the Laplace transform of the 
cosine $\int_0^\infty{\rm d} t\ e^{-\lambda t} \cos(\omega t) = \lambda/(\lambda^2+\omega^2)$ 
with Eqs.~(\ref{eq:gamma}) and~(\ref{app:gk}) we 
obtain 
\begin{align}\label{eq:gL}
\frac{1}{M}\tilde g_k(\lambda) = \frac{\lambda}{\nu_k^2-\lambda^2}\left(|\nu_k|\tilde\gamma(|\nu_k|) - 
\lambda\tilde\gamma(\lambda)\right) 
\; . 
\end{align}
By using instead Eqs.~(\ref{eq:gamma}) and~(\ref{eq:zeta}) we derive
\begin{eqnarray}
\zeta_k = |\nu_k|\tilde\gamma(|\nu_k|)
\; . 
\end{eqnarray}
The kernel $\tilde\gamma$ can be eliminated in favor of $\tilde{\mathcal G}_+$ through Eq.~(\ref{eq:Gplus}) which yields
\begin{eqnarray}\label{eq:hk}
\tilde h_k(\lambda) = \frac{\lambda}{\nu_k^2 - \lambda^2}\left[\tilde{\mathcal G}_+^{-1}(|\nu_k|) - \tilde{\mathcal G}_+^{-1}(\lambda)\right]\; .
\end{eqnarray}
This expression can now be inserted via Eq.~(\ref{eq:123}) into Eq.~(\ref{eq:corr2}) to find the equilibrium correlator. Note that, for 
$\omega_0 = \omega$ we have $\tilde{\mathcal G}_+(|\nu_k|) = u_k$. 
With the help of the one variable function
\begin{eqnarray}\label{eq:corrL}
\tilde {\mathcal C}^{\rm 1eq}(\lambda) 
= 
\frac{1}{\beta M} \sum_k \frac{\lambda}{\nu_k^2-\lambda^2}\left[\tilde{\mathcal G}_+(\lambda)-\tilde{\mathcal G}_+(|\nu_k|)\right]
\end{eqnarray}
 the equilibrium correlation function becomes 
\begin{eqnarray}\label{eq:corrLL}
\tilde {\mathcal C}^{\rm eq}(\lambda,\kappa) = \frac{{\mathcal C}^{\rm 1eq}(\lambda)+
{\mathcal C}^{\rm 1eq}(\kappa)}{\lambda + \kappa} 
\; ,
\end{eqnarray}
which clearly displays time translational invariance (TTI). Indeed, the Laplace transform with respect to $t$ and $t'$ of a generic 
function $f(|t-t'|)$ that depends only on the time difference is equal to $[\tilde f(\lambda) + \tilde f(\kappa)]/(\lambda + \kappa)$, 
where $\tilde f(\lambda)$ denotes the Laplace transform of $f(t)$ with respect to $t$. Hence, we have 
$\mathcal C^{\rm eq}(t,t') = \mathcal C^{\rm 1eq}(|t-t'|)$ with the explicit Laplace representation of $\mathcal C^{\rm 1eq}$ in 
Eq.~(\ref{eq:corrL}). The equilibrium correlation function is 
thus found without imposing TTI. 
By imposing TTI Eq.~(\ref{eq:corrL}) can be directly found from Eq.~(\ref{eq:corrgen}) by setting $t' = 0$ which simplifies the expression 
considerably. Remember that $\dot{\mathcal G}_+(t=0) = 1$ and $\mathcal G_+(t=0) = 0$. By taking the Laplace 
transform of the result 
with respect to $t$ and by using Eq.~(\ref{eq:gL}) one easily recovers Eq.~(\ref{eq:corrL}).

It is now straightforward to establish the relation between $\mathcal C^{\rm 1eq}(t)$ and $\mathcal R(t)$. Firstly, we note that since 
$\mathcal C^{\rm 1eq}(t)$ is an even function of $t$ its Fourier transform $\mathcal C^{\rm 1eq}(\omega)$ is related to its 
Laplace transform through
\begin{eqnarray}
\mathcal C^{\rm 1eq}(\omega) = \tilde {\mathcal C}^{\rm 1eq}(i\omega) + \tilde {\mathcal C}^{\rm 1eq}(-i\omega) \; .
\end{eqnarray}
Thus, by using Eqs.~(\ref{eq:chi}) and~(\ref{eq:corrL}) we have
\begin{align}
&{\mathcal C}^{\rm 1eq}(\omega) = \frac{1}{\beta}\sum_k \frac{i\omega}{\omega^2 + 
\nu_k^2}\left[\tilde{\mathcal R}(i\omega) - \tilde{\mathcal R}(-i\omega)\right]
\; .
\end{align}
Now, since the Fourier transform of the response function, ${\mathcal R}(\omega)$, is related  to its Laplace transform via 
$\mathcal R(\omega) = \tilde{\mathcal R}(i\omega)$ due to causality we obtain the quantum FDT in the form
\begin{eqnarray}
C^{\rm 1eq}(\omega) = \hbar\coth[\omega\beta\hbar/2] \im {\mathcal R}(\omega)
\;  ,
\end{eqnarray}
where we used the formula $\sum_k \omega/(\omega^2 + \nu_k^2) = (\beta\hbar/2) \coth[\omega\beta\hbar/2]$.
This result is completely general, in the sense that it applies to any bath, as it should.


\section{Asymptotic behavior of $\mathcal G_+(t)$ and $\mathcal C^{\rm 1eq}(t)$ for Ohmic dissipation}
\label{app:longtime}

In the case of \emph{Ohmic dissipation} the spectral function has the form 
\begin{eqnarray}
S(\omega) = \gamma\omega\;\;\;\mathrm{for}\;\;\;\omega \to 0 
\; .
\end{eqnarray}
For large frequencies one typically introduces a high frequency cutoff function (since the ultraviolet divergence is unphysical)
that we choose to be of the Drude--type $\omega_D^2/(\omega_D^2 + \omega^2)$ where $\omega_D \gg \omega$ is the high 
frequency cut-off. The bath kernel 
\begin{eqnarray}
\gamma(t) = \gamma\omega_D e^{-\omega_D t}
\end{eqnarray}
then has a finite memory and a simple form in the Laplace domain, namely
\begin{eqnarray}\label{gammaOhmic}
\tilde\gamma(\lambda) = \gamma\frac{\omega_D}{\omega_D + \lambda} 
\; .
\end{eqnarray}
We are interested in the equilibration behavior of the correlation function Eq.~(\ref{eq:corrinitialmeasure}) when quantum effects 
dominate. In order to find the long--time behavior of $\tilde {\mathcal C}^{\rm 1eq}(t)$ and $\mathcal G_+(t)$ we study their 
small--$\lambda$ behavior. In the very low temperature limit the sum over the Matsubara frequencies in Eq.~(\ref{eq:corrL}) can be 
replaced by an integral. For $\lambda\to 0$ one finds
\begin{align}
\tilde {\mathcal C}^{\rm 1eq}(\lambda) &\simeq 
\int_0^\infty \frac{{\rm d}\nu}{\pi}\frac{\lambda}{\nu^2 - \lambda^2}\left[\tilde{\mathcal G}_+(\lambda) - \tilde{\mathcal G}_+(\nu)\right]
\nonumber\\
&\simeq\frac{1}{\pi} \int_0^{\lambda^{-1}}\frac{{\rm d}\nu}{\nu^2 - 1} \ \frac{\lambda^2(\nu^2-1) + \lambda (\nu\tilde\gamma(\lambda\nu) - \gamma)}
{\omega^4} \label{eq:eq1}\\
&+ \frac{\lambda}{\pi}\int_1^\infty \frac{{\rm d}\nu}{\nu^2} \ \frac{\nu^2 + \nu\tilde\gamma(\nu)}{\omega^2(\nu^2 + \nu\tilde\gamma(\nu)+\omega^2)} +  ... \; ,
\nonumber
\end{align}
where the ellipsis stands for higher orders in $\lambda$. Now, by observing the ultraviolet behavior of Eq.~(\ref{gammaOhmic}) one easily 
argues that all the terms in the rhs of Eq.~(\ref{eq:eq1}) are of order $\sim \lambda$. Therefore, the long time behavior of the Ohmic 
equilibrium correlation function at zero temperature is
\begin{eqnarray}\label{eq:longC}
C^{\rm 1eq} \sim \frac{1}{t^2} \;\;\;\mathrm{for}\;\;\;t \to \infty 
\;\;\;\mathrm{and}\;\;\;\beta\hbar \gg |\omega^2 - \gamma^2/4|^{-1/2} \; .
\end{eqnarray}
It is straightforward to show by direct inversion of the Laplace transform that the propagator $\mathcal G^+(t)$ 
is \emph{exponentially} suppressed for large times (and for Ohmic dissipation) on a typical time scale $\gamma/2$, hence 
we have
\begin{eqnarray}\label{eq:longG}
\mathcal G_+(t) \sim e^{-\gamma t/2}\;\;\;\mathrm{for}\;\;\;t \to \infty \; .
\end{eqnarray}
Equation~(\ref{eq:longG}) holds for all temperatures. In the high temperature regime one has $\nu_k \to \infty$ so that 
$\mathcal C^{\rm 1eq} (\lambda)\simeq -\left[\tilde{\mathcal G}_+(\lambda)-1/\omega^2\right] / (\beta M \lambda)$. Translated into 
real time this states that $\mathcal G_+(t)$ is proportional to the derivative of $\mathcal C^{\rm 1eq}(t)$ which is nothing else than 
the classical FDT. Accordingly, we find
\begin{eqnarray}
\mathcal C^{\rm 1eq} \sim e^{-\gamma t/2} \;\;\;\mathrm{for}\;\;\;t \to \infty 
\;\;\;\mathrm{and}\;\;\;\beta\hbar \ll |\omega^2 - \gamma^2/4|^{-1/2} \; .
\end{eqnarray}

\end{appendices}

\acknowledgements
We thank T. Giamarchi for very useful discussions. 
This work was financially
supported by ANR-BLAN-0346 (FAMOUS).


\newpage

\bibliographystyle{phjcp}
\bibliography{artbib1}

\begin{thebibliography}{10}

\bibitem{We08}
{\sc U.~Weiss},
\newblock {\em Quantum Dissipative Systems},
\newblock World Scientific Publishing Co., Singapore, 2008.

\bibitem{SchZaik90}
{\sc G.~Sch{\"o}n} and {\sc A.~Zaikin},
\newblock {\em Phys. Rep.} {\bf 5}, 237 (1990).

\bibitem{CalLegg87}
{\sc A.~Leggett}, {\sc S.~Chakravarty}, {\sc A.~Dorsey}, {\sc M.~Fisher}, {\sc
  A.~Garg}, and {\sc W.~Zwerger},
\newblock {\em Rev. Mod. Phys.} {\bf 59}, 1 (1987).

\bibitem{Hanggi90}
{\sc P.~H\"anggi}, {\sc P.~Talkner}, and {\sc M.~Borkovec},
\newblock {\em Rev. Mod. Phys.} {\bf 62}, 251 (1990).

\bibitem{FeynVern63}
{\sc R.~Feynman} and {\sc F.~Vernon},
\newblock {\em Ann. Phys. (N.Y.)} {\bf 24}, 118 (1963).

\bibitem{Schwinger61}
{\sc J.~Schwinger},
\newblock {\em J. Math. Phys.} {\bf 2}, 407 (1961).

\bibitem{CalLegg83}
{\sc A.~Caldeira} and {\sc A.~Leggett},
\newblock {\em Physica A} {\bf 121}, 587 (1983).

\bibitem{Agarwal71}
{\sc G.~Agarwal},
\newblock {\em Phys. Rev. A} {\bf 4}, 739 (1971).

\bibitem{Grscin88}
{\sc H.~Grabert}, {\sc P.~Schramm}, and {\sc G.-L. Ingold},
\newblock {\em Phys. Rep.} {\bf 168}, 115 (1988).

\bibitem{Pottier00}
{\sc N.~Pottier} and {\sc A.~Mauger},
\newblock {\em Physica A} {\bf 282}, 77 (2000).

\bibitem{Ingold02}
{\sc G.~Ingold},
\newblock in {\em Coherent Evolution in Noisy Systems}, volume 611, p.~1,
  Springer-Verlag Berlin, 2002.

\bibitem{HuPaz92}
{\sc B.~Hu}, {\sc J.~Paz}, and {\sc Y.~Zhang},
\newblock {\em Phys. Rev. D} {\bf 45}, 2843 (1992).

\bibitem{Grab97}
{\sc R.~Karrlein} and {\sc H.~Grabert},
\newblock {\em Phys. Rev. E} {\bf 55}, 153 (1997).

\bibitem{Ford65}
{\sc G.~W. Ford}, {\sc M.~Kac}, and {\sc P.~Mazur},
\newblock {\em J. Math. Phys.} {\bf 6}, 504 (1965).

\bibitem{Gardiner04}
{\sc C.~W. Gardiner} and {\sc P.~Zoller},
\newblock {\em Quantum Noise},
\newblock Springer-Verlag, Berlin, 2004.

\bibitem{Giamarchi11}
{\sc J.~Catani}, {\sc G.~Lamporesi}, {\sc D.~Naik}, {\sc M.~Gring}, {\sc
  M.~Inguscio}, {\sc F.~Minardi}, {\sc A.~Kantian}, and {\sc T.~Giamarchi},
\newblock {\em Phys. Rev. A} {\bf 85}, 023623 (2012).

\bibitem{Giamarchi07}
{\sc M.~Zvonarev}, {\sc V.~Cheianov}, and {\sc T.~Giamarchi},
\newblock {\em Phys. Rev. Lett.} {\bf 99}, 240404 (2007).

\bibitem{Palzer09}
{\sc S.~Palzer}, {\sc C.~Zipkes}, {\sc C.~Sias}, and {\sc M.~K\"ohl},
\newblock {\em Phys. Rev. Lett.} {\bf 103}, 150601 (2009).

\bibitem{Johnson12}
{\sc T.~H. Johnson}, {\sc M.~Bruderer}, {\sc Y.~Cai}, {\sc S.~R. Clark}, {\sc
  W.~Bao}, and {\sc D.~Jaksch},
\newblock {\em Europhys. Lett.} {\bf 98}, 26001 (2012).

\bibitem{Calzetta02}
{\sc E.~Calzetta}, {\sc A.~Roura}, and {\sc E.~Verdaguer},
\newblock {\em Physica A} {\bf 319}, 188 (2003).

\bibitem{RoschKopp95}
{\sc A.~Rosch} and {\sc T.~Kopp},
\newblock {\em Phys. Rev. Lett.} {\bf 75}, 1988 (1995).

\bibitem{SchiroZwierlein09}
{\sc A.~Schirotzek}, {\sc C.-H. Wu}, {\sc A.~Sommer}, and {\sc M.~W.
  Zwierlein},
\newblock {\em Phys. Rev. Lett.} {\bf 102}, 230402 (2009).

\bibitem{Pita04}
{\sc G.~E. Astrakharchik} and {\sc L.~P. Pitaevskii},
\newblock {\em Phys. Rev. A} {\bf 70}, 013608 (2004).

\bibitem{Caux09}
{\sc A.~Y. Cherny}, {\sc J.-S. Caux}, and {\sc J.~Brand},
\newblock {\em Phys. Rev. A} {\bf 80}, 043604 (2009).

\bibitem{Devreese09}
{\sc J.~T. Devreese} and {\sc A.~S. Alexandrov},
\newblock {\em Reports on Progress in Physics} {\bf 72}, 066501 (2009).

\bibitem{Tempere09}
{\sc J.~Tempere}, {\sc W.~Casteels}, {\sc M.~K. Oberthaler}, {\sc S.~Knoop},
  {\sc E.~Timmermans}, and {\sc J.~T. Devreese},
\newblock {\em Physical Review B} {\bf 80}, 184504 (2009).

\bibitem{Hanggi05}
{\sc P.~H\"anggi} and {\sc G.~Ingold},
\newblock {\em Chaos} {\bf 15}, 026105 (2005).

\bibitem{Cugliandolo06}
{\sc L.~F. Cugliandolo}, {\sc T.~Giamarchi}, and {\sc P.~L. Doussal},
\newblock {\em Phys. Rev. Lett.} {\bf 96}, 217203 (2006).

\bibitem{Culo98}
{\sc L.~F. Cugliandolo} and {\sc G.~S. Lozano},
\newblock {\em Phys. Rev. Lett.} {\bf 80}, 4979 (1998).

\bibitem{Culo99}
{\sc L.~F. Cugliandolo} and {\sc G.~S. Lozano},
\newblock {\em Phys. Rev. B} {\bf 59}, 915 (1999).

\bibitem{Cugrlolosa}
{\sc L.~F. Cugliandolo}, {\sc D.~R. Grempel}, {\sc G.~S. Lozano}, {\sc
  H.~Lozza}, and {\sc C.~A. {da Silva Santos}},
\newblock {\em Phys. Rev. B} {\bf 66}, 014444 (2002).

\bibitem{Kech}
{\sc M.~P. Kennett} and {\sc C.~Chamon},
\newblock {\em Phys. Rev. Lett.} {\bf 86}, 1622 (2001).

\bibitem{Kechye}
{\sc M.~P. Kennett}, {\sc C.~Chamon}, and {\sc Y.~Ye},
\newblock {\em Phys. Rev. B} {\bf 64}, 224408 (2001).

\bibitem{Bipa}
{\sc G.~Biroli} and {\sc O.~Parcollet},
\newblock {\em Phys. Rev. B} {\bf 65}, 094414 (2002).

\bibitem{Arbicu10}
{\sc C.~Aron}, {\sc G.~Biroli}, and {\sc L.~F. Cugliandolo},
\newblock {\em J. Stat. Mech.} , P11018 (2010).

\bibitem{Arbicu10b}
{\sc C.~Aron}, {\sc G.~Biroli}, and {\sc L.~F. Cugliandolo},
\newblock {\em Phys. Rev. B} {\bf 82}, 174203 (2010).

\bibitem{Tomo50}
{\sc S.~Tomonaga},
\newblock {\em Prog. in Theor. Phys.} {\bf 5}, 544 (1950).

\bibitem{Lutt63}
{\sc J.~M. Luttinger},
\newblock {\em J. Math. Phys.} {\bf 4}, 1154 (1963).

\bibitem{MattisLieb65}
{\sc D.~Mattis} and {\sc E.~Lieb},
\newblock {\em J. Math. Phys.} {\bf 6}, 304 (1965).

\bibitem{Thal08}
{\sc G.~Thalhammer}, {\sc G.~Barontini}, {\sc L.~D. Sarlo}, {\sc J.~Catani},
  {\sc F.~Minardi}, and {\sc M.~Inguscio},
\newblock {\em Phys. Rev. Lett.} {\bf 100}, 210402 (2008).

\bibitem{Peano05}
{\sc V.~Peano}, {\sc M.~Thorwart}, {\sc C.~Mora}, and {\sc R.~Egger},
\newblock {\em New J. Phys.} {\bf 7}, 192 (2005).

\bibitem{Olsh98}
{\sc M.~Olshanii},
\newblock {\em Phys. Rev. Lett.} {\bf 81}, 938 (1998).

\bibitem{Lieb-Liniger}
{\sc E.~H. Lieb} and {\sc W.~Liniger},
\newblock {\em Phys. Rev.} {\bf 130}, 1605Ð1616 (1963).

\bibitem{Giamarchi03}
{\sc T.~Giamarchi},
\newblock {\em Quantum Physics in One Dimension},
\newblock Clarendon Press, Oxford, 2003.

\bibitem{Note1}
The underlying probability distribution is of the (Gaussian) Boltzmann-Gibbs
  type $\protect \qopname \relax o{exp}(-\beta \protect \mathcal H)$ with
  $\protect \mathcal H$ the \protect \emph {full coupled} Hamiltonian of the
  particle--bath system. Equation~(\ref {langevin}) thus describes the case
  where the harmonic oscillator bath and the particle are initially \protect
  \emph {coupled} as in the quantum case studied in the present work. This
  subtle point is often overlooked. For more details see p.~21-23 in \cite
  {We08}.

\bibitem{Bocuga11}
{\sc J.~Bonart}, {\sc L.~F. Cugliandolo}, and {\sc A.~Gambassi},
\newblock {\em J. Stat. Mech.} , P01014 (2012).

\end{thebibliography}

\end{document}